\documentclass[a4paper]{report}

\usepackage{amsmath}
\usepackage{amssymb}
\usepackage{url}
\usepackage{colortbl}
\usepackage{graphicx}

\newcommand{\plus}{\ensuremath{\raisebox{0.1em}{$\scriptscriptstyle\mathord{+}$}\negthickspace}}
\renewcommand{\star}{\ensuremath{\raisebox{0.1em}{$\scriptstyle\mathord{*}$}\!}}
\newcommand{\opt}{\ensuremath{\mathord{?}}\!}

\author{Vadim Zaytsev, \href{mailto:vadim@grammarware.net}{\url{vadim@grammarware.net}},\\
Software Analysis \& Transformation Team (SWAT),\\Centrum Wiskunde \& Informatica (CWI), The Netherlands}
\title{{\Huge Guided Grammar Convergence}\\Full Case Study Report\\
\Large Generated by \texttt{converge::Guided}}

\usepackage[unicode,bookmarks=false,pdfstartview={FitH},%
            colorlinks,linkcolor=blue,urlcolor=blue,citecolor=blue,%
            pdfauthor={Dr. Vadim Zaytsev},
            pdftitle={Guided Grammar Convergence: Full Case Study Report}]{hyperref}
\begin{document}
\maketitle

\chapter*{Introduction}

\begin{figure}[b!] % htbp
	\centering
		\includegraphics[width=\textwidth]{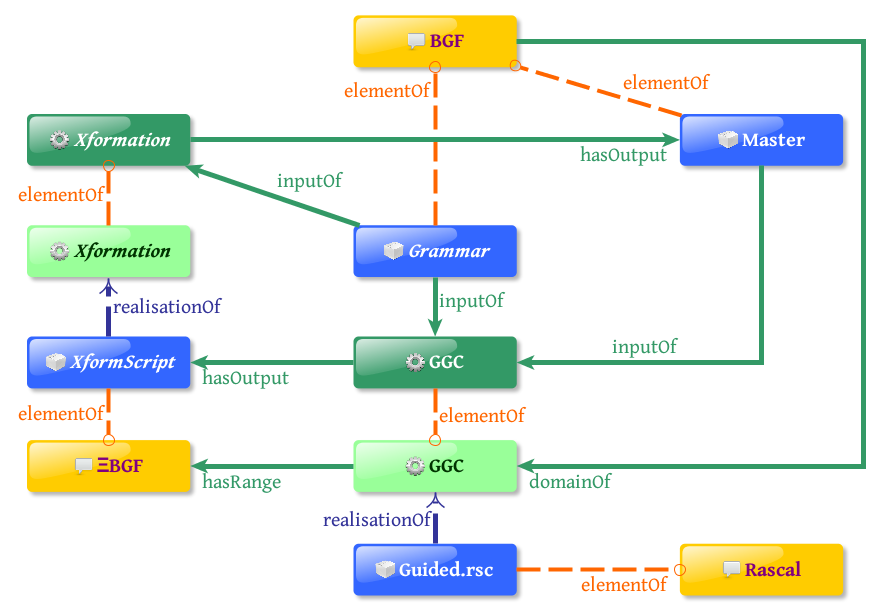}
	\caption{Guided grammar convergence megamodel.}
	\label{mega}
\end{figure}

This report is meant to be used as auxiliary material for the \emph{guided grammar convergence} technique proposed in
\cite{Guided2013} as problem-specific improvement on \cite{Convergence2009}. It contains a megamodel renarrated as
proposed in \cite{Renarration2012}, as well as full results of the guided grammar convergence experiment on the
Factorial Language, with details about each grammar source packaged in a readable form. All formulae used within this
document, are generated automatically by the convergence infrastructure in order to avoid any mistakes. The generator
source code and the source of the introduction text can be found publicly available in the Software Language Processing
Suite repository~\cite{SLPS}.

Consider the model on \autoref{mega}. It is a \emph{megamodel} in the sense of \cite{BJV04,FNG04}, since it depicts a
\emph{linguistic architecture}: all nodes represent software languages and language transformations, and all edges
represent relationships between them. MegaL~\cite{MegaL} is used as a notation: blue boxes represent tangible
\emph{artefacts} (files, programs, modules, directories, collections of other concrete entities), yellow boxes denote
software \emph{languages} in the broad sense (from general purpose programming languages to data types and protocols),
light green boxes are used for \emph{functions} (in fact, model transformations) and dark green boxes are for
\emph{function applications}.

As we can see from \autoref{mega} if we start reading it from the bottom, there is a program
\href{https://github.com/grammarware/slps/blob/master/shared/rascal/src/converge/Guided.rsc}{Guided.rsc}, which was
written in Rascal metaprogramming language~\cite{Rascal}. It implements the guided grammar convergence process, which
input language is BGF (BNF-like Grammar Formalism, a straightforward internal representation format for grammars,
introduced in \cite{Convergence2009}). Its output language is $\Xi$BGF, a bidirectional grammar transformation language
introduced in \cite{Metasyntactically2012}. An application of the guided grammar convergence algorithm to two grammars:
one \emph{master grammar} defining the \emph{intended} software language (terminology of \cite{Guided2013}) and one
servant grammar (its label displayed in italics since it is actually a variable, not a concrete entity) --- yields a
transformation script that implements a grammar transformation than indeed transforms the servant grammar into the
master grammar. The process behind this inference is relatively complicated and involves triggered grammar design
mutations, normalisation to Abstract Normal Form, constructing weak prodsig-equivalence ($\Bumpeq$) classes and
resolving nominal and structural differences, as described on the theoretic level in \cite{Guided2013}.

The rest of the report presents instantiations of this megamodel for eleven concrete grammar sources:

\begin{description}
\item[adt:] an algebraic data type\footnote{\url{http://tutor.rascal-mpl.org/Courses/Rascal/Declarations/AlgebraicDataType/AlgebraicDataType.html}.} in Rascal~\cite{RascalTutor};
\item[antlr:] a parser description in the input language of ANTLR~\cite{ANTLR}.
Semantic actions (in Java) are intertwined with EBNF-like productions.
\item[dcg:] a logic program written in the style of definite clause grammars~\cite{DCG}.
% \item[ecore:] an Ecore model~\cite{MOF}, created manually in Eclipse modeling framework~\cite{EMF}
% 	and represented in XMI;
\item[emf:] an %alternative 
	Ecore model~\cite{MOF}, automatically generated by Eclipse~\cite{EMF} from the XML Schema of the
	\textbf{xsd} source;
\item[jaxb:] an object model obtained by a data binding framework.
Generated automatically by JAXB~\cite{JSR31} from the XML schema for FL.
\item[om:] a hand-crafted object model (Java classes) for the abstract syntax of FL.
\item[python:] a parser specification in a scripting language, using the PyParsing library~\cite{McGuire2007};
\item[rascal:] a concrete syntax specification in Rascal metaprogramming language \cite{RascalTutor,Rascal};
\item[sdf:] a concrete syntax definition in the notation of SDF~\cite{Klint93} with scannerless
generalized LR~\cite{EKV09,Visser97} as a parsing model.
\item[txl:] a concrete syntax definition in the notation of TXL (Turing eXtender Language) transformational framework \cite{DeanCMS02}, which, unlike SDF, uses a combination of pattern matching and term rewriting).
\item[xsd:] an XML schema~\cite{W3C-XSD} for the abstract syntax of FL.
\end{description}

\tableofcontents

\chapter{ANTLR}

 Source name: \textbf{antlr}

\section{Source grammar}

\begin{itemize}\item Source artifact: \href{http://github.com/grammarware/slps/blob/master/topics/fl/java1/FL.g}{topics/fl/java1/FL.g}\item Grammar extractor: \href{http://github.com/grammarware/slps/blob/master/topics/extraction/antlr/antlrstrip.py}{topics/extraction/antlr/antlrstrip.py}\item Grammar extractor: \href{http://github.com/grammarware/slps/blob/master/topics/extraction/antlr/slps/antlr2bgf/StrippedANTLR.g}{topics/extraction/antlr/slps/antlr2bgf/StrippedANTLR.g}\end{itemize}

\footnotesize\begin{center}\begin{tabular}{|l|}\hline
\multicolumn{1}{|>{\columncolor[gray]{.9}}c|}{\footnotesize \textbf{Production rules}}
\\\hline
$\mathrm{p}(\text{`'},\mathit{program},\plus \left(\mathrm{sel}\left(\text{`f'},\mathit{function}\right)\right))$	\\
$\mathrm{p}(\text{`'},\mathit{function},\mathrm{seq}\left(\left[\mathrm{sel}\left(\text{`n'},\mathit{ID}\right), \plus \left(\mathrm{sel}\left(\text{`a'},\mathit{ID}\right)\right), \text{`='}, \mathrm{sel}\left(\text{`e'},\mathit{expr}\right), \plus \left(\mathit{NEWLINE}\right)\right]\right))$	\\
$\mathrm{p}(\text{`'},\mathit{expr},\mathrm{choice}([\mathrm{sel}\left(\text{`b'},\mathit{binary}\right),$\\$\qquad\qquad\mathrm{sel}\left(\text{`a'},\mathit{apply}\right),$\\$\qquad\qquad\mathrm{sel}\left(\text{`i'},\mathit{ifThenElse}\right)]))$	\\
$\mathrm{p}(\text{`'},\mathit{binary},\mathrm{seq}\left(\left[\mathrm{sel}\left(\text{`l'},\mathit{atom}\right), \star \left(\mathrm{seq}\left(\left[\mathrm{sel}\left(\text{`o'},\mathit{ops}\right), \mathrm{sel}\left(\text{`r'},\mathit{atom}\right)\right]\right)\right)\right]\right))$	\\
$\mathrm{p}(\text{`'},\mathit{apply},\mathrm{seq}\left(\left[\mathrm{sel}\left(\text{`i'},\mathit{ID}\right), \plus \left(\mathrm{sel}\left(\text{`a'},\mathit{atom}\right)\right)\right]\right))$	\\
$\mathrm{p}(\text{`'},\mathit{ifThenElse},\mathrm{seq}\left(\left[\text{`if'}, \mathrm{sel}\left(\text{`c'},\mathit{expr}\right), \text{`then'}, \mathrm{sel}\left(\text{`e1'},\mathit{expr}\right), \text{`else'}, \mathrm{sel}\left(\text{`e2'},\mathit{expr}\right)\right]\right))$	\\
$\mathrm{p}(\text{`'},\mathit{atom},\mathrm{choice}([\mathit{ID},$\\$\qquad\qquad\mathit{INT},$\\$\qquad\qquad\mathrm{seq}\left(\left[\text{`('}, \mathrm{sel}\left(\text{`e'},\mathit{expr}\right), \text{`)'}\right]\right)]))$	\\
$\mathrm{p}(\text{`'},\mathit{ops},\mathrm{choice}([\text{`=='},$\\$\qquad\qquad\text{`+'},$\\$\qquad\qquad\text{`-'}]))$	\\
\hline\end{tabular}\end{center}

\section{Mutations}
{\footnotesize\begin{itemize}
\item \textbf{unite-splitN} $expr$ \\$\mathrm{p}\left(\text{`'},\mathit{atom},\mathrm{choice}\left(\left[\mathit{ID}, \mathit{INT}, \mathrm{seq}\left(\left[\text{`('}, \mathrm{sel}\left(\text{`e'},\mathit{expr}\right), \text{`)'}\right]\right)\right]\right)\right)$
\item \textbf{designate-unlabel}\\$\mathrm{p}\left(\fbox{\text{`tmplabel'}},\mathit{binary},\mathrm{seq}\left(\left[\mathrm{sel}\left(\text{`l'},\mathit{expr}\right), \star \left(\mathrm{seq}\left(\left[\mathrm{sel}\left(\text{`o'},\mathit{ops}\right), \mathrm{sel}\left(\text{`r'},\mathit{expr}\right)\right]\right)\right)\right]\right)\right)$
\item \textbf{anonymize-deanonymize}\\$\mathrm{p}\left(\text{`tmplabel'},\mathit{binary},\mathrm{seq}\left(\left[\fbox{$\mathrm{sel}\left(\text{`l'},\mathit{expr}\right)$}, \star \left(\mathrm{seq}\left(\left[\fbox{$\mathrm{sel}\left(\text{`o'},\mathit{ops}\right)$}, \fbox{$\mathrm{sel}\left(\text{`r'},\mathit{expr}\right)$}\right]\right)\right)\right]\right)\right)$
\item \textbf{assoc-iterate}\\$\mathrm{p}\left(\text{`tmplabel'},\mathit{binary},\mathrm{seq}\left(\left[\mathit{expr}, \mathit{ops}, \mathit{expr}\right]\right)\right)$
\item \textbf{deanonymize-anonymize}\\$\mathrm{p}\left(\text{`tmplabel'},\mathit{binary},\mathrm{seq}\left(\left[\fbox{$\mathrm{sel}\left(\text{`l'},\mathit{expr}\right)$}, \fbox{$\mathrm{sel}\left(\text{`o'},\mathit{ops}\right)$}, \fbox{$\mathrm{sel}\left(\text{`r'},\mathit{expr}\right)$}\right]\right)\right)$
\item \textbf{unlabel-designate}\\$\mathrm{p}\left(\fbox{\text{`tmplabel'}},\mathit{binary},\mathrm{seq}\left(\left[\mathrm{sel}\left(\text{`l'},\mathit{expr}\right), \mathrm{sel}\left(\text{`o'},\mathit{ops}\right), \mathrm{sel}\left(\text{`r'},\mathit{expr}\right)\right]\right)\right)$
\end{itemize}}

\section{Normalizations}
{\footnotesize\begin{itemize}
\item \textbf{reroot-reroot} $\left[\right]$ to $\left[\mathit{program}\right]$
\item \textbf{anonymize-deanonymize}\\$\mathrm{p}\left(\text{`'},\mathit{function},\mathrm{seq}\left(\left[\fbox{$\mathrm{sel}\left(\text{`n'},\mathit{ID}\right)$}, \plus \left(\fbox{$\mathrm{sel}\left(\text{`a'},\mathit{ID}\right)$}\right), \text{`='}, \fbox{$\mathrm{sel}\left(\text{`e'},\mathit{expr}\right)$}, \plus \left(\mathit{NEWLINE}\right)\right]\right)\right)$
\item \textbf{anonymize-deanonymize}\\$\mathrm{p}\left(\text{`'},\mathit{program},\plus \left(\fbox{$\mathrm{sel}\left(\text{`f'},\mathit{function}\right)$}\right)\right)$
\item \textbf{anonymize-deanonymize}\\$\mathrm{p}\left(\text{`'},\mathit{ifThenElse},\mathrm{seq}\left(\left[\text{`if'}, \fbox{$\mathrm{sel}\left(\text{`c'},\mathit{expr}\right)$}, \text{`then'}, \fbox{$\mathrm{sel}\left(\text{`e1'},\mathit{expr}\right)$}, \text{`else'}, \fbox{$\mathrm{sel}\left(\text{`e2'},\mathit{expr}\right)$}\right]\right)\right)$
\item \textbf{anonymize-deanonymize}\\$\mathrm{p}\left(\text{`'},\mathit{binary},\mathrm{seq}\left(\left[\fbox{$\mathrm{sel}\left(\text{`l'},\mathit{expr}\right)$}, \fbox{$\mathrm{sel}\left(\text{`o'},\mathit{ops}\right)$}, \fbox{$\mathrm{sel}\left(\text{`r'},\mathit{expr}\right)$}\right]\right)\right)$
\item \textbf{anonymize-deanonymize}\\$\mathrm{p}\left(\text{`'},\mathit{expr},\mathrm{choice}\left(\left[\mathit{ID}, \mathit{INT}, \mathrm{seq}\left(\left[\text{`('}, \fbox{$\mathrm{sel}\left(\text{`e'},\mathit{expr}\right)$}, \text{`)'}\right]\right)\right]\right)\right)$
\item \textbf{anonymize-deanonymize}\\$\mathrm{p}\left(\text{`'},\mathit{apply},\mathrm{seq}\left(\left[\fbox{$\mathrm{sel}\left(\text{`i'},\mathit{ID}\right)$}, \plus \left(\fbox{$\mathrm{sel}\left(\text{`a'},\mathit{expr}\right)$}\right)\right]\right)\right)$
\item \textbf{anonymize-deanonymize}\\$\mathrm{p}\left(\text{`'},\mathit{expr},\mathrm{choice}\left(\left[\fbox{$\mathrm{sel}\left(\text{`b'},\mathit{binary}\right)$}, \fbox{$\mathrm{sel}\left(\text{`a'},\mathit{apply}\right)$}, \fbox{$\mathrm{sel}\left(\text{`i'},\mathit{ifThenElse}\right)$}\right]\right)\right)$
\item \textbf{abstractize-concretize}\\$\mathrm{p}\left(\text{`'},\mathit{ops},\mathrm{choice}\left(\left[\fbox{$\text{`=='}$}, \fbox{$\text{`+'}$}, \fbox{$\text{`-'}$}\right]\right)\right)$
\item \textbf{abstractize-concretize}\\$\mathrm{p}\left(\text{`'},\mathit{expr},\mathrm{choice}\left(\left[\mathit{ID}, \mathit{INT}, \mathrm{seq}\left(\left[\fbox{$\text{`('}$}, \mathit{expr}, \fbox{$\text{`)'}$}\right]\right)\right]\right)\right)$
\item \textbf{abstractize-concretize}\\$\mathrm{p}\left(\text{`'},\mathit{function},\mathrm{seq}\left(\left[\mathit{ID}, \plus \left(\mathit{ID}\right), \fbox{$\text{`='}$}, \mathit{expr}, \plus \left(\mathit{NEWLINE}\right)\right]\right)\right)$
\item \textbf{abstractize-concretize}\\$\mathrm{p}\left(\text{`'},\mathit{ifThenElse},\mathrm{seq}\left(\left[\fbox{$\text{`if'}$}, \mathit{expr}, \fbox{$\text{`then'}$}, \mathit{expr}, \fbox{$\text{`else'}$}, \mathit{expr}\right]\right)\right)$
\item \textbf{vertical-horizontal}  in $\mathit{expr}$
\item \textbf{undefine-define}\\$\mathrm{p}\left(\text{`'},\mathit{ops},\varepsilon\right)$
\item \textbf{unchain-chain}\\$\mathrm{p}\left(\text{`'},\mathit{expr},\mathit{binary}\right)$
\item \textbf{unchain-chain}\\$\mathrm{p}\left(\text{`'},\mathit{expr},\mathit{apply}\right)$
\item \textbf{unchain-chain}\\$\mathrm{p}\left(\text{`'},\mathit{expr},\mathit{ifThenElse}\right)$
\item \textbf{abridge-detour}\\$\mathrm{p}\left(\text{`'},\mathit{expr},\mathit{expr}\right)$
\item \textbf{unlabel-designate}\\$\mathrm{p}\left(\fbox{\text{`binary'}},\mathit{expr},\mathrm{seq}\left(\left[\mathit{expr}, \mathit{ops}, \mathit{expr}\right]\right)\right)$
\item \textbf{unlabel-designate}\\$\mathrm{p}\left(\fbox{\text{`apply'}},\mathit{expr},\mathrm{seq}\left(\left[\mathit{ID}, \plus \left(\mathit{expr}\right)\right]\right)\right)$
\item \textbf{unlabel-designate}\\$\mathrm{p}\left(\fbox{\text{`ifThenElse'}},\mathit{expr},\mathrm{seq}\left(\left[\mathit{expr}, \mathit{expr}, \mathit{expr}\right]\right)\right)$
\item \textbf{extract-inline}  in $\mathit{expr}$\\$\mathrm{p}\left(\text{`'},\mathit{expr_1},\mathrm{seq}\left(\left[\mathit{expr}, \mathit{ops}, \mathit{expr}\right]\right)\right)$
\item \textbf{extract-inline}  in $\mathit{expr}$\\$\mathrm{p}\left(\text{`'},\mathit{expr_2},\mathrm{seq}\left(\left[\mathit{ID}, \plus \left(\mathit{expr}\right)\right]\right)\right)$
\item \textbf{extract-inline}  in $\mathit{expr}$\\$\mathrm{p}\left(\text{`'},\mathit{expr_3},\mathrm{seq}\left(\left[\mathit{expr}, \mathit{expr}, \mathit{expr}\right]\right)\right)$
\end{itemize}}

\section{Grammar in ANF}

\footnotesize\begin{center}\begin{tabular}{|l|c|}\hline
\multicolumn{1}{|>{\columncolor[gray]{.9}}c|}{\footnotesize \textbf{Production rule}} &
\multicolumn{1}{>{\columncolor[gray]{.9}}c|}{\footnotesize \textbf{Production signature}}
\\\hline
$\mathrm{p}\left(\text{`'},\mathit{program},\plus \left(\mathit{function}\right)\right)$	&	$\{ \langle \mathit{function}, {+}\rangle\}$\\
$\mathrm{p}\left(\text{`'},\mathit{function},\mathrm{seq}\left(\left[\mathit{ID}, \plus \left(\mathit{ID}\right), \mathit{expr}, \plus \left(\mathit{NEWLINE}\right)\right]\right)\right)$	&	$\{ \langle \mathit{expr}, 1\rangle, \langle \mathit{NEWLINE}, {+}\rangle, \langle \mathit{ID}, 1{+}\rangle\}$\\
$\mathrm{p}\left(\text{`'},\mathit{expr},\mathit{ID}\right)$	&	$\{ \langle \mathit{ID}, 1\rangle\}$\\
$\mathrm{p}\left(\text{`'},\mathit{expr},\mathit{INT}\right)$	&	$\{ \langle \mathit{INT}, 1\rangle\}$\\
$\mathrm{p}\left(\text{`'},\mathit{expr},\mathit{expr_1}\right)$	&	$\{ \langle \mathit{expr_1}, 1\rangle\}$\\
$\mathrm{p}\left(\text{`'},\mathit{expr},\mathit{expr_2}\right)$	&	$\{ \langle \mathit{expr_2}, 1\rangle\}$\\
$\mathrm{p}\left(\text{`'},\mathit{expr},\mathit{expr_3}\right)$	&	$\{ \langle \mathit{expr_3}, 1\rangle\}$\\
$\mathrm{p}\left(\text{`'},\mathit{expr_1},\mathrm{seq}\left(\left[\mathit{expr}, \mathit{ops}, \mathit{expr}\right]\right)\right)$	&	$\{ \langle \mathit{ops}, 1\rangle, \langle \mathit{expr}, 11\rangle\}$\\
$\mathrm{p}\left(\text{`'},\mathit{expr_2},\mathrm{seq}\left(\left[\mathit{ID}, \plus \left(\mathit{expr}\right)\right]\right)\right)$	&	$\{ \langle \mathit{expr}, {+}\rangle, \langle \mathit{ID}, 1\rangle\}$\\
$\mathrm{p}\left(\text{`'},\mathit{expr_3},\mathrm{seq}\left(\left[\mathit{expr}, \mathit{expr}, \mathit{expr}\right]\right)\right)$	&	$\{ \langle \mathit{expr}, 111\rangle\}$\\
\hline\end{tabular}\end{center}

\section{Nominal resolution}

Production rules are matched as follows (ANF on the left, master grammar on the right):
\begin{eqnarray*}
\mathrm{p}\left(\text{`'},\mathit{program},\plus \left(\mathit{function}\right)\right) & \bumpeq & \mathrm{p}\left(\text{`'},\mathit{program},\plus \left(\mathit{function}\right)\right) \\
\mathrm{p}\left(\text{`'},\mathit{function},\mathrm{seq}\left(\left[\mathit{ID}, \plus \left(\mathit{ID}\right), \mathit{expr}, \plus \left(\mathit{NEWLINE}\right)\right]\right)\right) & \Bumpeq & \mathrm{p}\left(\text{`'},\mathit{function},\mathrm{seq}\left(\left[str, \plus \left(str\right), \mathit{expression}\right]\right)\right) \\
\mathrm{p}\left(\text{`'},\mathit{expr},\mathit{ID}\right) & \bumpeq & \mathrm{p}\left(\text{`'},\mathit{expression},str\right) \\
\mathrm{p}\left(\text{`'},\mathit{expr},\mathit{INT}\right) & \bumpeq & \mathrm{p}\left(\text{`'},\mathit{expression},int\right) \\
\mathrm{p}\left(\text{`'},\mathit{expr},\mathit{expr_1}\right) & \bumpeq & \mathrm{p}\left(\text{`'},\mathit{expression},\mathit{binary}\right) \\
\mathrm{p}\left(\text{`'},\mathit{expr},\mathit{expr_2}\right) & \bumpeq & \mathrm{p}\left(\text{`'},\mathit{expression},\mathit{apply}\right) \\
\mathrm{p}\left(\text{`'},\mathit{expr},\mathit{expr_3}\right) & \bumpeq & \mathrm{p}\left(\text{`'},\mathit{expression},\mathit{conditional}\right) \\
\mathrm{p}\left(\text{`'},\mathit{expr_1},\mathrm{seq}\left(\left[\mathit{expr}, \mathit{ops}, \mathit{expr}\right]\right)\right) & \bumpeq & \mathrm{p}\left(\text{`'},\mathit{binary},\mathrm{seq}\left(\left[\mathit{expression}, \mathit{operator}, \mathit{expression}\right]\right)\right) \\
\mathrm{p}\left(\text{`'},\mathit{expr_2},\mathrm{seq}\left(\left[\mathit{ID}, \plus \left(\mathit{expr}\right)\right]\right)\right) & \bumpeq & \mathrm{p}\left(\text{`'},\mathit{apply},\mathrm{seq}\left(\left[str, \plus \left(\mathit{expression}\right)\right]\right)\right) \\
\mathrm{p}\left(\text{`'},\mathit{expr_3},\mathrm{seq}\left(\left[\mathit{expr}, \mathit{expr}, \mathit{expr}\right]\right)\right) & \bumpeq & \mathrm{p}\left(\text{`'},\mathit{conditional},\mathrm{seq}\left(\left[\mathit{expression}, \mathit{expression}, \mathit{expression}\right]\right)\right) \\
\end{eqnarray*}
This yields the following nominal mapping:
\begin{align*}\mathit{antlr} \:\diamond\: \mathit{master} =\:& \{\langle \mathit{program},\mathit{program}\rangle,\\
 & \langle \mathit{expr_3},\mathit{conditional}\rangle,\\
 & \langle \mathit{expr_1},\mathit{binary}\rangle,\\
 & \langle \mathit{function},\mathit{function}\rangle,\\
 & \langle \mathit{ID},str\rangle,\\
 & \langle \mathit{expr},\mathit{expression}\rangle,\\
 & \langle \mathit{INT},int\rangle,\\
 & \langle \mathit{ops},\mathit{operator}\rangle,\\
 & \langle \mathit{NEWLINE},\omega\rangle,\\
 & \langle \mathit{expr_2},\mathit{apply}\rangle\}\end{align*}
 Which is exercised with these grammar transformation steps:

{\footnotesize\begin{itemize}
\item \textbf{renameN-renameN} $\mathit{expr_3}$ to $\mathit{conditional}$
\item \textbf{renameN-renameN} $\mathit{expr_1}$ to $\mathit{binary}$
\item \textbf{renameN-renameN} $\mathit{ID}$ to $str$
\item \textbf{renameN-renameN} $\mathit{expr}$ to $\mathit{expression}$
\item \textbf{renameN-renameN} $\mathit{INT}$ to $int$
\item \textbf{renameN-renameN} $\mathit{ops}$ to $\mathit{operator}$
\item \textbf{renameN-renameN} $\mathit{expr_2}$ to $\mathit{apply}$
\end{itemize}}

\section{Structural resolution}
{\footnotesize\begin{itemize}
\item \textbf{project-inject}\\$\mathrm{p}\left(\text{`'},\mathit{function},\mathrm{seq}\left(\left[str, \plus \left(str\right), \mathit{expression}, \plus \left(\fbox{$\mathit{NEWLINE}$}\right)\right]\right)\right)$
\end{itemize}}

\chapter{Definite Clause Grammar}

 Source name: \textbf{dcg}

\section{Source grammar}

\begin{itemize}\item Source artifact: \href{http://github.com/grammarware/slps/blob/master/topics/fl/prolog1/Parser.pro}{topics/fl/prolog1/Parser.pro}\item Grammar extractor: \href{http://github.com/grammarware/slps/blob/master/shared/prolog/cli/dcg2bgf.pro}{shared/prolog/cli/dcg2bgf.pro}\end{itemize}

\footnotesize\begin{center}\begin{tabular}{|l|}\hline
\multicolumn{1}{|>{\columncolor[gray]{.9}}c|}{\footnotesize \textbf{Production rules}}
\\\hline
$\mathrm{p}(\text{`'},\mathit{program},\plus \left(\mathit{function}\right))$	\\
$\mathrm{p}(\text{`'},\mathit{function},\mathrm{seq}\left(\left[\mathit{name}, \plus \left(\mathit{name}\right), \text{`='}, \mathit{expr}, \plus \left(\mathit{newline}\right)\right]\right))$	\\
$\mathrm{p}(\text{`binary'},\mathit{expr},\mathrm{seq}\left(\left[\mathit{atom}, \star \left(\mathrm{seq}\left(\left[\mathit{ops}, \mathit{atom}\right]\right)\right)\right]\right))$	\\
$\mathrm{p}(\text{`apply'},\mathit{expr},\mathrm{seq}\left(\left[\mathit{name}, \plus \left(\mathit{atom}\right)\right]\right))$	\\
$\mathrm{p}(\text{`ifThenElse'},\mathit{expr},\mathrm{seq}\left(\left[\text{`if'}, \mathit{expr}, \text{`then'}, \mathit{expr}, \text{`else'}, \mathit{expr}\right]\right))$	\\
$\mathrm{p}(\text{`literal'},\mathit{atom},\mathit{int})$	\\
$\mathrm{p}(\text{`argument'},\mathit{atom},\mathit{name})$	\\
$\mathrm{p}(\text{`'},\mathit{atom},\mathrm{seq}\left(\left[\text{`('}, \mathit{expr}, \text{`)'}\right]\right))$	\\
$\mathrm{p}(\text{`equal'},\mathit{ops},\text{`=='})$	\\
$\mathrm{p}(\text{`plus'},\mathit{ops},\text{`+'})$	\\
$\mathrm{p}(\text{`minus'},\mathit{ops},\text{`-'})$	\\
\hline\end{tabular}\end{center}

\section{Mutations}
{\footnotesize\begin{itemize}
\item \textbf{unite-splitN} $expr$ \\$\mathrm{p}\left(\text{`literal'},\mathit{atom},\mathit{int}\right)$\\$\mathrm{p}\left(\text{`argument'},\mathit{atom},\mathit{name}\right)$\\$\mathrm{p}\left(\text{`'},\mathit{atom},\mathrm{seq}\left(\left[\text{`('}, \mathit{expr}, \text{`)'}\right]\right)\right)$
\item \textbf{assoc-iterate}\\$\mathrm{p}\left(\text{`binary'},\mathit{expr},\mathrm{seq}\left(\left[\mathit{expr}, \mathit{ops}, \mathit{expr}\right]\right)\right)$
\end{itemize}}

\section{Normalizations}
{\footnotesize\begin{itemize}
\item \textbf{reroot-reroot} $\left[\right]$ to $\left[\mathit{program}\right]$
\item \textbf{unlabel-designate}\\$\mathrm{p}\left(\fbox{\text{`binary'}},\mathit{expr},\mathrm{seq}\left(\left[\mathit{expr}, \mathit{ops}, \mathit{expr}\right]\right)\right)$
\item \textbf{unlabel-designate}\\$\mathrm{p}\left(\fbox{\text{`apply'}},\mathit{expr},\mathrm{seq}\left(\left[\mathit{name}, \plus \left(\mathit{expr}\right)\right]\right)\right)$
\item \textbf{unlabel-designate}\\$\mathrm{p}\left(\fbox{\text{`ifThenElse'}},\mathit{expr},\mathrm{seq}\left(\left[\text{`if'}, \mathit{expr}, \text{`then'}, \mathit{expr}, \text{`else'}, \mathit{expr}\right]\right)\right)$
\item \textbf{unlabel-designate}\\$\mathrm{p}\left(\fbox{\text{`literal'}},\mathit{expr},\mathit{int}\right)$
\item \textbf{unlabel-designate}\\$\mathrm{p}\left(\fbox{\text{`argument'}},\mathit{expr},\mathit{name}\right)$
\item \textbf{unlabel-designate}\\$\mathrm{p}\left(\fbox{\text{`equal'}},\mathit{ops},\text{`=='}\right)$
\item \textbf{unlabel-designate}\\$\mathrm{p}\left(\fbox{\text{`plus'}},\mathit{ops},\text{`+'}\right)$
\item \textbf{unlabel-designate}\\$\mathrm{p}\left(\fbox{\text{`minus'}},\mathit{ops},\text{`-'}\right)$
\item \textbf{abstractize-concretize}\\$\mathrm{p}\left(\text{`'},\mathit{expr},\mathrm{seq}\left(\left[\fbox{$\text{`('}$}, \mathit{expr}, \fbox{$\text{`)'}$}\right]\right)\right)$
\item \textbf{abstractize-concretize}\\$\mathrm{p}\left(\text{`'},\mathit{function},\mathrm{seq}\left(\left[\mathit{name}, \plus \left(\mathit{name}\right), \fbox{$\text{`='}$}, \mathit{expr}, \plus \left(\mathit{newline}\right)\right]\right)\right)$
\item \textbf{abstractize-concretize}\\$\mathrm{p}\left(\text{`'},\mathit{ops},\fbox{$\text{`=='}$}\right)$
\item \textbf{abstractize-concretize}\\$\mathrm{p}\left(\text{`'},\mathit{expr},\mathrm{seq}\left(\left[\fbox{$\text{`if'}$}, \mathit{expr}, \fbox{$\text{`then'}$}, \mathit{expr}, \fbox{$\text{`else'}$}, \mathit{expr}\right]\right)\right)$
\item \textbf{abstractize-concretize}\\$\mathrm{p}\left(\text{`'},\mathit{ops},\fbox{$\text{`-'}$}\right)$
\item \textbf{abstractize-concretize}\\$\mathrm{p}\left(\text{`'},\mathit{ops},\fbox{$\text{`+'}$}\right)$
\item \textbf{undefine-define}\\$\mathrm{p}\left(\text{`'},\mathit{ops},\varepsilon\right)$
\item \textbf{abridge-detour}\\$\mathrm{p}\left(\text{`'},\mathit{expr},\mathit{expr}\right)$
\item \textbf{extract-inline}  in $\mathit{expr}$\\$\mathrm{p}\left(\text{`'},\mathit{expr_1},\mathrm{seq}\left(\left[\mathit{expr}, \mathit{ops}, \mathit{expr}\right]\right)\right)$
\item \textbf{extract-inline}  in $\mathit{expr}$\\$\mathrm{p}\left(\text{`'},\mathit{expr_2},\mathrm{seq}\left(\left[\mathit{name}, \plus \left(\mathit{expr}\right)\right]\right)\right)$
\item \textbf{extract-inline}  in $\mathit{expr}$\\$\mathrm{p}\left(\text{`'},\mathit{expr_3},\mathrm{seq}\left(\left[\mathit{expr}, \mathit{expr}, \mathit{expr}\right]\right)\right)$
\end{itemize}}

\section{Grammar in ANF}

\footnotesize\begin{center}\begin{tabular}{|l|c|}\hline
\multicolumn{1}{|>{\columncolor[gray]{.9}}c|}{\footnotesize \textbf{Production rule}} &
\multicolumn{1}{>{\columncolor[gray]{.9}}c|}{\footnotesize \textbf{Production signature}}
\\\hline
$\mathrm{p}\left(\text{`'},\mathit{program},\plus \left(\mathit{function}\right)\right)$	&	$\{ \langle \mathit{function}, {+}\rangle\}$\\
$\mathrm{p}\left(\text{`'},\mathit{function},\mathrm{seq}\left(\left[\mathit{name}, \plus \left(\mathit{name}\right), \mathit{expr}, \plus \left(\mathit{newline}\right)\right]\right)\right)$	&	$\{ \langle \mathit{expr}, 1\rangle, \langle \mathit{newline}, {+}\rangle, \langle \mathit{name}, 1{+}\rangle\}$\\
$\mathrm{p}\left(\text{`'},\mathit{expr},\mathit{expr_1}\right)$	&	$\{ \langle \mathit{expr_1}, 1\rangle\}$\\
$\mathrm{p}\left(\text{`'},\mathit{expr},\mathit{expr_2}\right)$	&	$\{ \langle \mathit{expr_2}, 1\rangle\}$\\
$\mathrm{p}\left(\text{`'},\mathit{expr},\mathit{expr_3}\right)$	&	$\{ \langle \mathit{expr_3}, 1\rangle\}$\\
$\mathrm{p}\left(\text{`'},\mathit{expr},\mathit{int}\right)$	&	$\{ \langle \mathit{int}, 1\rangle\}$\\
$\mathrm{p}\left(\text{`'},\mathit{expr},\mathit{name}\right)$	&	$\{ \langle \mathit{name}, 1\rangle\}$\\
$\mathrm{p}\left(\text{`'},\mathit{expr_1},\mathrm{seq}\left(\left[\mathit{expr}, \mathit{ops}, \mathit{expr}\right]\right)\right)$	&	$\{ \langle \mathit{ops}, 1\rangle, \langle \mathit{expr}, 11\rangle\}$\\
$\mathrm{p}\left(\text{`'},\mathit{expr_2},\mathrm{seq}\left(\left[\mathit{name}, \plus \left(\mathit{expr}\right)\right]\right)\right)$	&	$\{ \langle \mathit{expr}, {+}\rangle, \langle \mathit{name}, 1\rangle\}$\\
$\mathrm{p}\left(\text{`'},\mathit{expr_3},\mathrm{seq}\left(\left[\mathit{expr}, \mathit{expr}, \mathit{expr}\right]\right)\right)$	&	$\{ \langle \mathit{expr}, 111\rangle\}$\\
\hline\end{tabular}\end{center}

\section{Nominal resolution}

Production rules are matched as follows (ANF on the left, master grammar on the right):
\begin{eqnarray*}
\mathrm{p}\left(\text{`'},\mathit{program},\plus \left(\mathit{function}\right)\right) & \bumpeq & \mathrm{p}\left(\text{`'},\mathit{program},\plus \left(\mathit{function}\right)\right) \\
\mathrm{p}\left(\text{`'},\mathit{function},\mathrm{seq}\left(\left[\mathit{name}, \plus \left(\mathit{name}\right), \mathit{expr}, \plus \left(\mathit{newline}\right)\right]\right)\right) & \Bumpeq & \mathrm{p}\left(\text{`'},\mathit{function},\mathrm{seq}\left(\left[str, \plus \left(str\right), \mathit{expression}\right]\right)\right) \\
\mathrm{p}\left(\text{`'},\mathit{expr},\mathit{expr_1}\right) & \bumpeq & \mathrm{p}\left(\text{`'},\mathit{expression},\mathit{binary}\right) \\
\mathrm{p}\left(\text{`'},\mathit{expr},\mathit{expr_2}\right) & \bumpeq & \mathrm{p}\left(\text{`'},\mathit{expression},\mathit{apply}\right) \\
\mathrm{p}\left(\text{`'},\mathit{expr},\mathit{expr_3}\right) & \bumpeq & \mathrm{p}\left(\text{`'},\mathit{expression},\mathit{conditional}\right) \\
\mathrm{p}\left(\text{`'},\mathit{expr},\mathit{int}\right) & \bumpeq & \mathrm{p}\left(\text{`'},\mathit{expression},int\right) \\
\mathrm{p}\left(\text{`'},\mathit{expr},\mathit{name}\right) & \bumpeq & \mathrm{p}\left(\text{`'},\mathit{expression},str\right) \\
\mathrm{p}\left(\text{`'},\mathit{expr_1},\mathrm{seq}\left(\left[\mathit{expr}, \mathit{ops}, \mathit{expr}\right]\right)\right) & \bumpeq & \mathrm{p}\left(\text{`'},\mathit{binary},\mathrm{seq}\left(\left[\mathit{expression}, \mathit{operator}, \mathit{expression}\right]\right)\right) \\
\mathrm{p}\left(\text{`'},\mathit{expr_2},\mathrm{seq}\left(\left[\mathit{name}, \plus \left(\mathit{expr}\right)\right]\right)\right) & \bumpeq & \mathrm{p}\left(\text{`'},\mathit{apply},\mathrm{seq}\left(\left[str, \plus \left(\mathit{expression}\right)\right]\right)\right) \\
\mathrm{p}\left(\text{`'},\mathit{expr_3},\mathrm{seq}\left(\left[\mathit{expr}, \mathit{expr}, \mathit{expr}\right]\right)\right) & \bumpeq & \mathrm{p}\left(\text{`'},\mathit{conditional},\mathrm{seq}\left(\left[\mathit{expression}, \mathit{expression}, \mathit{expression}\right]\right)\right) \\
\end{eqnarray*}
This yields the following nominal mapping:
\begin{align*}\mathit{dcg} \:\diamond\: \mathit{master} =\:& \{\langle \mathit{program},\mathit{program}\rangle,\\
 & \langle \mathit{expr_3},\mathit{conditional}\rangle,\\
 & \langle \mathit{expr_1},\mathit{binary}\rangle,\\
 & \langle \mathit{function},\mathit{function}\rangle,\\
 & \langle \mathit{expr},\mathit{expression}\rangle,\\
 & \langle \mathit{name},str\rangle,\\
 & \langle \mathit{ops},\mathit{operator}\rangle,\\
 & \langle \mathit{int},int\rangle,\\
 & \langle \mathit{newline},\omega\rangle,\\
 & \langle \mathit{expr_2},\mathit{apply}\rangle\}\end{align*}
 Which is exercised with these grammar transformation steps:

{\footnotesize\begin{itemize}
\item \textbf{renameN-renameN} $\mathit{expr_3}$ to $\mathit{conditional}$
\item \textbf{renameN-renameN} $\mathit{expr_1}$ to $\mathit{binary}$
\item \textbf{renameN-renameN} $\mathit{expr}$ to $\mathit{expression}$
\item \textbf{renameN-renameN} $\mathit{name}$ to $str$
\item \textbf{renameN-renameN} $\mathit{ops}$ to $\mathit{operator}$
\item \textbf{renameN-renameN} $\mathit{int}$ to $int$
\item \textbf{renameN-renameN} $\mathit{expr_2}$ to $\mathit{apply}$
\end{itemize}}

\section{Structural resolution}
{\footnotesize\begin{itemize}
\item \textbf{project-inject}\\$\mathrm{p}\left(\text{`'},\mathit{function},\mathrm{seq}\left(\left[str, \plus \left(str\right), \mathit{expression}, \plus \left(\fbox{$\mathit{newline}$}\right)\right]\right)\right)$
\end{itemize}}

\chapter{Eclipse Modeling Framework}

 Source name: \textbf{emf}

\section{Source grammar}

\begin{itemize}\item Source artifact: \href{http://github.com/grammarware/slps/blob/master/topics/fl/emf2/model/fl.ecore}{topics/fl/emf2/model/fl.ecore}\item Grammar extractor: \href{http://github.com/grammarware/slps/blob/master/topics/extraction/ecore/ecore2bgf.xslt}{topics/extraction/ecore/ecore2bgf.xslt}\end{itemize}

\footnotesize\begin{center}\begin{tabular}{|l|}\hline
\multicolumn{1}{|>{\columncolor[gray]{.9}}c|}{\footnotesize \textbf{Production rules}}
\\\hline
$\mathrm{p}(\text{`'},\mathit{Apply},\mathrm{seq}\left(\left[\mathrm{sel}\left(\text{`name'},str\right), \plus \left(\mathrm{sel}\left(\text{`arg'},\mathit{Expr}\right)\right)\right]\right))$	\\
$\mathrm{p}(\text{`'},\mathit{Argument},\mathrm{sel}\left(\text{`name'},str\right))$	\\
$\mathrm{p}(\text{`'},\mathit{Binary},\mathrm{seq}\left(\left[\mathrm{sel}\left(\text{`ops'},\mathit{Ops}\right), \mathrm{sel}\left(\text{`left'},\mathit{Expr}\right), \mathrm{sel}\left(\text{`right'},\mathit{Expr}\right)\right]\right))$	\\
$\mathrm{p}(\text{`'},\mathit{Expr},\mathrm{choice}([\mathit{Apply},$\\$\qquad\qquad\mathit{Argument},$\\$\qquad\qquad\mathit{Binary},$\\$\qquad\qquad\mathit{IfThenElse},$\\$\qquad\qquad\mathit{Literal}]))$	\\
$\mathrm{p}(\text{`'},\mathit{Function},\mathrm{seq}\left(\left[\mathrm{sel}\left(\text{`name'},str\right), \plus \left(\mathrm{sel}\left(\text{`arg'},str\right)\right), \mathrm{sel}\left(\text{`rhs'},\mathit{Expr}\right)\right]\right))$	\\
$\mathrm{p}(\text{`'},\mathit{IfThenElse},\mathrm{seq}\left(\left[\mathrm{sel}\left(\text{`ifExpr'},\mathit{Expr}\right), \mathrm{sel}\left(\text{`thenExpr'},\mathit{Expr}\right), \mathrm{sel}\left(\text{`elseExpr'},\mathit{Expr}\right)\right]\right))$	\\
$\mathrm{p}(\text{`'},\mathit{Literal},\mathrm{sel}\left(\text{`info'},int\right))$	\\
$\mathrm{p}(\text{`'},\mathit{Ops},\mathrm{choice}([\mathrm{sel}\left(\text{`Equal'},\varepsilon\right),$\\$\qquad\qquad\mathrm{sel}\left(\text{`Plus'},\varepsilon\right),$\\$\qquad\qquad\mathrm{sel}\left(\text{`Minus'},\varepsilon\right)]))$	\\
$\mathrm{p}(\text{`'},\mathit{ProgramType},\plus \left(\mathrm{sel}\left(\text{`function'},\mathit{Function}\right)\right))$	\\
\hline\end{tabular}\end{center}

\section{Normalizations}
{\footnotesize\begin{itemize}
\item \textbf{reroot-reroot} $\left[\right]$ to $\left[\mathit{ProgramType}\right]$
\item \textbf{unlabel-designate}\\$\mathrm{p}\left(\fbox{\text{`name'}},\mathit{Argument},str\right)$
\item \textbf{unlabel-designate}\\$\mathrm{p}\left(\fbox{\text{`info'}},\mathit{Literal},int\right)$
\item \textbf{anonymize-deanonymize}\\$\mathrm{p}\left(\text{`'},\mathit{Function},\mathrm{seq}\left(\left[\fbox{$\mathrm{sel}\left(\text{`name'},str\right)$}, \plus \left(\fbox{$\mathrm{sel}\left(\text{`arg'},str\right)$}\right), \fbox{$\mathrm{sel}\left(\text{`rhs'},\mathit{Expr}\right)$}\right]\right)\right)$
\item \textbf{anonymize-deanonymize}\\$\mathrm{p}\left(\text{`'},\mathit{Apply},\mathrm{seq}\left(\left[\fbox{$\mathrm{sel}\left(\text{`name'},str\right)$}, \plus \left(\fbox{$\mathrm{sel}\left(\text{`arg'},\mathit{Expr}\right)$}\right)\right]\right)\right)$
\item \textbf{anonymize-deanonymize}\\$\mathrm{p}\left(\text{`'},\mathit{IfThenElse},\mathrm{seq}\left(\left[\fbox{$\mathrm{sel}\left(\text{`ifExpr'},\mathit{Expr}\right)$}, \fbox{$\mathrm{sel}\left(\text{`thenExpr'},\mathit{Expr}\right)$}, \fbox{$\mathrm{sel}\left(\text{`elseExpr'},\mathit{Expr}\right)$}\right]\right)\right)$
\item \textbf{anonymize-deanonymize}\\$\mathrm{p}\left(\text{`'},\mathit{Ops},\mathrm{choice}\left(\left[\fbox{$\mathrm{sel}\left(\text{`Equal'},\varepsilon\right)$}, \fbox{$\mathrm{sel}\left(\text{`Plus'},\varepsilon\right)$}, \fbox{$\mathrm{sel}\left(\text{`Minus'},\varepsilon\right)$}\right]\right)\right)$
\item \textbf{anonymize-deanonymize}\\$\mathrm{p}\left(\text{`'},\mathit{ProgramType},\plus \left(\fbox{$\mathrm{sel}\left(\text{`function'},\mathit{Function}\right)$}\right)\right)$
\item \textbf{anonymize-deanonymize}\\$\mathrm{p}\left(\text{`'},\mathit{Binary},\mathrm{seq}\left(\left[\fbox{$\mathrm{sel}\left(\text{`ops'},\mathit{Ops}\right)$}, \fbox{$\mathrm{sel}\left(\text{`left'},\mathit{Expr}\right)$}, \fbox{$\mathrm{sel}\left(\text{`right'},\mathit{Expr}\right)$}\right]\right)\right)$
\item \textbf{vertical-horizontal}  in $\mathit{Expr}$
\item \textbf{undefine-define}\\$\mathrm{p}\left(\text{`'},\mathit{Ops},\varepsilon\right)$
\item \textbf{unchain-chain}\\$\mathrm{p}\left(\text{`'},\mathit{Expr},\mathit{Apply}\right)$
\item \textbf{unchain-chain}\\$\mathrm{p}\left(\text{`'},\mathit{Expr},\mathit{Argument}\right)$
\item \textbf{unchain-chain}\\$\mathrm{p}\left(\text{`'},\mathit{Expr},\mathit{Binary}\right)$
\item \textbf{unchain-chain}\\$\mathrm{p}\left(\text{`'},\mathit{Expr},\mathit{IfThenElse}\right)$
\item \textbf{unchain-chain}\\$\mathrm{p}\left(\text{`'},\mathit{Expr},\mathit{Literal}\right)$
\item \textbf{unlabel-designate}\\$\mathrm{p}\left(\fbox{\text{`Apply'}},\mathit{Expr},\mathrm{seq}\left(\left[str, \plus \left(\mathit{Expr}\right)\right]\right)\right)$
\item \textbf{unlabel-designate}\\$\mathrm{p}\left(\fbox{\text{`Argument'}},\mathit{Expr},str\right)$
\item \textbf{unlabel-designate}\\$\mathrm{p}\left(\fbox{\text{`Binary'}},\mathit{Expr},\mathrm{seq}\left(\left[\mathit{Ops}, \mathit{Expr}, \mathit{Expr}\right]\right)\right)$
\item \textbf{unlabel-designate}\\$\mathrm{p}\left(\fbox{\text{`IfThenElse'}},\mathit{Expr},\mathrm{seq}\left(\left[\mathit{Expr}, \mathit{Expr}, \mathit{Expr}\right]\right)\right)$
\item \textbf{unlabel-designate}\\$\mathrm{p}\left(\fbox{\text{`Literal'}},\mathit{Expr},int\right)$
\item \textbf{extract-inline}  in $\mathit{Expr}$\\$\mathrm{p}\left(\text{`'},\mathit{Expr_1},\mathrm{seq}\left(\left[str, \plus \left(\mathit{Expr}\right)\right]\right)\right)$
\item \textbf{extract-inline}  in $\mathit{Expr}$\\$\mathrm{p}\left(\text{`'},\mathit{Expr_2},\mathrm{seq}\left(\left[\mathit{Ops}, \mathit{Expr}, \mathit{Expr}\right]\right)\right)$
\item \textbf{extract-inline}  in $\mathit{Expr}$\\$\mathrm{p}\left(\text{`'},\mathit{Expr_3},\mathrm{seq}\left(\left[\mathit{Expr}, \mathit{Expr}, \mathit{Expr}\right]\right)\right)$
\end{itemize}}

\section{Grammar in ANF}

\footnotesize\begin{center}\begin{tabular}{|l|c|}\hline
\multicolumn{1}{|>{\columncolor[gray]{.9}}c|}{\footnotesize \textbf{Production rule}} &
\multicolumn{1}{>{\columncolor[gray]{.9}}c|}{\footnotesize \textbf{Production signature}}
\\\hline
$\mathrm{p}\left(\text{`'},\mathit{Expr},\mathit{Expr_1}\right)$	&	$\{ \langle \mathit{Expr_1}, 1\rangle\}$\\
$\mathrm{p}\left(\text{`'},\mathit{Expr},str\right)$	&	$\{ \langle str, 1\rangle\}$\\
$\mathrm{p}\left(\text{`'},\mathit{Expr},\mathit{Expr_2}\right)$	&	$\{ \langle \mathit{Expr_2}, 1\rangle\}$\\
$\mathrm{p}\left(\text{`'},\mathit{Expr},\mathit{Expr_3}\right)$	&	$\{ \langle \mathit{Expr_3}, 1\rangle\}$\\
$\mathrm{p}\left(\text{`'},\mathit{Expr},int\right)$	&	$\{ \langle int, 1\rangle\}$\\
$\mathrm{p}\left(\text{`'},\mathit{Function},\mathrm{seq}\left(\left[str, \plus \left(str\right), \mathit{Expr}\right]\right)\right)$	&	$\{ \langle str, 1{+}\rangle, \langle \mathit{Expr}, 1\rangle\}$\\
$\mathrm{p}\left(\text{`'},\mathit{ProgramType},\plus \left(\mathit{Function}\right)\right)$	&	$\{ \langle \mathit{Function}, {+}\rangle\}$\\
$\mathrm{p}\left(\text{`'},\mathit{Expr_1},\mathrm{seq}\left(\left[str, \plus \left(\mathit{Expr}\right)\right]\right)\right)$	&	$\{ \langle str, 1\rangle, \langle \mathit{Expr}, {+}\rangle\}$\\
$\mathrm{p}\left(\text{`'},\mathit{Expr_2},\mathrm{seq}\left(\left[\mathit{Ops}, \mathit{Expr}, \mathit{Expr}\right]\right)\right)$	&	$\{ \langle \mathit{Ops}, 1\rangle, \langle \mathit{Expr}, 11\rangle\}$\\
$\mathrm{p}\left(\text{`'},\mathit{Expr_3},\mathrm{seq}\left(\left[\mathit{Expr}, \mathit{Expr}, \mathit{Expr}\right]\right)\right)$	&	$\{ \langle \mathit{Expr}, 111\rangle\}$\\
\hline\end{tabular}\end{center}

\section{Nominal resolution}

Production rules are matched as follows (ANF on the left, master grammar on the right):
\begin{eqnarray*}
\mathrm{p}\left(\text{`'},\mathit{Expr},\mathit{Expr_1}\right) & \bumpeq & \mathrm{p}\left(\text{`'},\mathit{expression},\mathit{apply}\right) \\
\mathrm{p}\left(\text{`'},\mathit{Expr},str\right) & \bumpeq & \mathrm{p}\left(\text{`'},\mathit{expression},str\right) \\
\mathrm{p}\left(\text{`'},\mathit{Expr},\mathit{Expr_2}\right) & \bumpeq & \mathrm{p}\left(\text{`'},\mathit{expression},\mathit{binary}\right) \\
\mathrm{p}\left(\text{`'},\mathit{Expr},\mathit{Expr_3}\right) & \bumpeq & \mathrm{p}\left(\text{`'},\mathit{expression},\mathit{conditional}\right) \\
\mathrm{p}\left(\text{`'},\mathit{Expr},int\right) & \bumpeq & \mathrm{p}\left(\text{`'},\mathit{expression},int\right) \\
\mathrm{p}\left(\text{`'},\mathit{Function},\mathrm{seq}\left(\left[str, \plus \left(str\right), \mathit{Expr}\right]\right)\right) & \bumpeq & \mathrm{p}\left(\text{`'},\mathit{function},\mathrm{seq}\left(\left[str, \plus \left(str\right), \mathit{expression}\right]\right)\right) \\
\mathrm{p}\left(\text{`'},\mathit{ProgramType},\plus \left(\mathit{Function}\right)\right) & \bumpeq & \mathrm{p}\left(\text{`'},\mathit{program},\plus \left(\mathit{function}\right)\right) \\
\mathrm{p}\left(\text{`'},\mathit{Expr_1},\mathrm{seq}\left(\left[str, \plus \left(\mathit{Expr}\right)\right]\right)\right) & \bumpeq & \mathrm{p}\left(\text{`'},\mathit{apply},\mathrm{seq}\left(\left[str, \plus \left(\mathit{expression}\right)\right]\right)\right) \\
\mathrm{p}\left(\text{`'},\mathit{Expr_2},\mathrm{seq}\left(\left[\mathit{Ops}, \mathit{Expr}, \mathit{Expr}\right]\right)\right) & \Bumpeq & \mathrm{p}\left(\text{`'},\mathit{binary},\mathrm{seq}\left(\left[\mathit{expression}, \mathit{operator}, \mathit{expression}\right]\right)\right) \\
\mathrm{p}\left(\text{`'},\mathit{Expr_3},\mathrm{seq}\left(\left[\mathit{Expr}, \mathit{Expr}, \mathit{Expr}\right]\right)\right) & \bumpeq & \mathrm{p}\left(\text{`'},\mathit{conditional},\mathrm{seq}\left(\left[\mathit{expression}, \mathit{expression}, \mathit{expression}\right]\right)\right) \\
\end{eqnarray*}
This yields the following nominal mapping:
\begin{align*}\mathit{emf} \:\diamond\: \mathit{master} =\:& \{\langle \mathit{Expr_2},\mathit{binary}\rangle,\\
 & \langle \mathit{ProgramType},\mathit{program}\rangle,\\
 & \langle \mathit{Expr_3},\mathit{conditional}\rangle,\\
 & \langle str,str\rangle,\\
 & \langle int,int\rangle,\\
 & \langle \mathit{Function},\mathit{function}\rangle,\\
 & \langle \mathit{Expr},\mathit{expression}\rangle,\\
 & \langle \mathit{Expr_1},\mathit{apply}\rangle,\\
 & \langle \mathit{Ops},\mathit{operator}\rangle\}\end{align*}
 Which is exercised with these grammar transformation steps:

{\footnotesize\begin{itemize}
\item \textbf{renameN-renameN} $\mathit{Expr_2}$ to $\mathit{binary}$
\item \textbf{renameN-renameN} $\mathit{ProgramType}$ to $\mathit{program}$
\item \textbf{renameN-renameN} $\mathit{Expr_3}$ to $\mathit{conditional}$
\item \textbf{renameN-renameN} $\mathit{Function}$ to $\mathit{function}$
\item \textbf{renameN-renameN} $\mathit{Expr}$ to $\mathit{expression}$
\item \textbf{renameN-renameN} $\mathit{Expr_1}$ to $\mathit{apply}$
\item \textbf{renameN-renameN} $\mathit{Ops}$ to $\mathit{operator}$
\end{itemize}}

\section{Structural resolution}
{\footnotesize\begin{itemize}
\item \textbf{permute-permute}\\$\mathrm{p}\left(\text{`'},\mathit{binary},\mathrm{seq}\left(\left[\mathit{operator}, \mathit{expression}, \mathit{expression}\right]\right)\right)$\\$\mathrm{p}\left(\text{`'},\mathit{binary},\mathrm{seq}\left(\left[\mathit{expression}, \mathit{operator}, \mathit{expression}\right]\right)\right)$
\end{itemize}}

\chapter{JAXB Data Binding Framework}

 Source name: \textbf{jaxb}

\section{Source grammar}

\begin{itemize}\item Source artifact: \href{http://github.com/grammarware/slps/blob/master/topics/fl/java3/fl/Apply.java}{topics/fl/java3/fl/Apply.java}\item Source artifact: \href{http://github.com/grammarware/slps/blob/master/topics/fl/java3/fl/Argument.java}{topics/fl/java3/fl/Argument.java}\item Source artifact: \href{http://github.com/grammarware/slps/blob/master/topics/fl/java3/fl/Binary.java}{topics/fl/java3/fl/Binary.java}\item Source artifact: \href{http://github.com/grammarware/slps/blob/master/topics/fl/java3/fl/Expr.java}{topics/fl/java3/fl/Expr.java}\item Source artifact: \href{http://github.com/grammarware/slps/blob/master/topics/fl/java3/fl/Function.java}{topics/fl/java3/fl/Function.java}\item Source artifact: \href{http://github.com/grammarware/slps/blob/master/topics/fl/java3/fl/IfThenElse.java}{topics/fl/java3/fl/IfThenElse.java}\item Source artifact: \href{http://github.com/grammarware/slps/blob/master/topics/fl/java3/fl/Literal.java}{topics/fl/java3/fl/Literal.java}\item Source artifact: \href{http://github.com/grammarware/slps/blob/master/topics/fl/java3/fl/ObjectFactory.java}{topics/fl/java3/fl/ObjectFactory.java}\item Source artifact: \href{http://github.com/grammarware/slps/blob/master/topics/fl/java3/fl/Ops.java}{topics/fl/java3/fl/Ops.java}\item Source artifact: \href{http://github.com/grammarware/slps/blob/master/topics/fl/java3/fl/Program.java}{topics/fl/java3/fl/Program.java}\item Source artifact: \href{http://github.com/grammarware/slps/blob/master/topics/fl/java3/fl/package-info.java}{topics/fl/java3/fl/package-info.java}\item Grammar extractor: \href{http://github.com/grammarware/slps/blob/master/topics/extraction/java2bgf/slps/java2bgf/Tool.java}{topics/extraction/java2bgf/slps/java2bgf/Tool.java}\end{itemize}

\footnotesize\begin{center}\begin{tabular}{|l|}\hline
\multicolumn{1}{|>{\columncolor[gray]{.9}}c|}{\footnotesize \textbf{Production rules}}
\\\hline
$\mathrm{p}(\text{`'},\mathit{Apply},\mathrm{seq}\left(\left[\mathrm{sel}\left(\text{`Name'},str\right), \mathrm{sel}\left(\text{`Arg'},\star \left(\mathit{Expr}\right)\right)\right]\right))$	\\
$\mathrm{p}(\text{`'},\mathit{Argument},\mathrm{sel}\left(\text{`Name'},str\right))$	\\
$\mathrm{p}(\text{`'},\mathit{Binary},\mathrm{seq}\left(\left[\mathrm{sel}\left(\text{`Ops'},\mathit{Ops}\right), \mathrm{sel}\left(\text{`Left'},\mathit{Expr}\right), \mathrm{sel}\left(\text{`Right'},\mathit{Expr}\right)\right]\right))$	\\
$\mathrm{p}(\text{`'},\mathit{Expr},\mathrm{choice}([\mathit{Apply},$\\$\qquad\qquad\mathit{Argument},$\\$\qquad\qquad\mathit{Binary},$\\$\qquad\qquad\mathit{IfThenElse},$\\$\qquad\qquad\mathit{Literal}]))$	\\
$\mathrm{p}(\text{`'},\mathit{Function},\mathrm{seq}\left(\left[\mathrm{sel}\left(\text{`Name'},str\right), \mathrm{sel}\left(\text{`Arg'},\star \left(str\right)\right), \mathrm{sel}\left(\text{`Rhs'},\mathit{Expr}\right)\right]\right))$	\\
$\mathrm{p}(\text{`'},\mathit{IfThenElse},\mathrm{seq}\left(\left[\mathrm{sel}\left(\text{`IfExpr'},\mathit{Expr}\right), \mathrm{sel}\left(\text{`ThenExpr'},\mathit{Expr}\right), \mathrm{sel}\left(\text{`ElseExpr'},\mathit{Expr}\right)\right]\right))$	\\
$\mathrm{p}(\text{`'},\mathit{Literal},\mathrm{sel}\left(\text{`Info'},int\right))$	\\
$\mathrm{p}(\text{`'},\mathit{ObjectFactory},\varepsilon)$	\\
$\mathrm{p}(\text{`'},\mathit{Ops},\mathrm{choice}([\mathrm{sel}\left(\text{`EQUAL'},\varepsilon\right),$\\$\qquad\qquad\mathrm{sel}\left(\text{`PLUS'},\varepsilon\right),$\\$\qquad\qquad\mathrm{sel}\left(\text{`MINUS'},\varepsilon\right)]))$	\\
$\mathrm{p}(\text{`'},\mathit{package-info},\varphi)$	\\
$\mathrm{p}(\text{`'},\mathit{Program},\mathrm{sel}\left(\text{`Function'},\star \left(\mathit{Function}\right)\right))$	\\
\hline\end{tabular}\end{center}

\section{Normalizations}
{\footnotesize\begin{itemize}
\item \textbf{reroot-reroot} $\left[\right]$ to $\left[\mathit{Program}\right]$
\item \textbf{unlabel-designate}\\$\mathrm{p}\left(\fbox{\text{`Name'}},\mathit{Argument},str\right)$
\item \textbf{unlabel-designate}\\$\mathrm{p}\left(\fbox{\text{`Info'}},\mathit{Literal},int\right)$
\item \textbf{unlabel-designate}\\$\mathrm{p}\left(\fbox{\text{`Function'}},\mathit{Program},\star \left(\mathit{Function}\right)\right)$
\item \textbf{anonymize-deanonymize}\\$\mathrm{p}\left(\text{`'},\mathit{IfThenElse},\mathrm{seq}\left(\left[\fbox{$\mathrm{sel}\left(\text{`IfExpr'},\mathit{Expr}\right)$}, \fbox{$\mathrm{sel}\left(\text{`ThenExpr'},\mathit{Expr}\right)$}, \fbox{$\mathrm{sel}\left(\text{`ElseExpr'},\mathit{Expr}\right)$}\right]\right)\right)$
\item \textbf{anonymize-deanonymize}\\$\mathrm{p}\left(\text{`'},\mathit{Function},\mathrm{seq}\left(\left[\fbox{$\mathrm{sel}\left(\text{`Name'},str\right)$}, \fbox{$\mathrm{sel}\left(\text{`Arg'},\star \left(str\right)\right)$}, \fbox{$\mathrm{sel}\left(\text{`Rhs'},\mathit{Expr}\right)$}\right]\right)\right)$
\item \textbf{anonymize-deanonymize}\\$\mathrm{p}\left(\text{`'},\mathit{Binary},\mathrm{seq}\left(\left[\fbox{$\mathrm{sel}\left(\text{`Ops'},\mathit{Ops}\right)$}, \fbox{$\mathrm{sel}\left(\text{`Left'},\mathit{Expr}\right)$}, \fbox{$\mathrm{sel}\left(\text{`Right'},\mathit{Expr}\right)$}\right]\right)\right)$
\item \textbf{anonymize-deanonymize}\\$\mathrm{p}\left(\text{`'},\mathit{Apply},\mathrm{seq}\left(\left[\fbox{$\mathrm{sel}\left(\text{`Name'},str\right)$}, \fbox{$\mathrm{sel}\left(\text{`Arg'},\star \left(\mathit{Expr}\right)\right)$}\right]\right)\right)$
\item \textbf{anonymize-deanonymize}\\$\mathrm{p}\left(\text{`'},\mathit{Ops},\mathrm{choice}\left(\left[\fbox{$\mathrm{sel}\left(\text{`EQUAL'},\varepsilon\right)$}, \fbox{$\mathrm{sel}\left(\text{`PLUS'},\varepsilon\right)$}, \fbox{$\mathrm{sel}\left(\text{`MINUS'},\varepsilon\right)$}\right]\right)\right)$
\item \textbf{vertical-horizontal}  in $\mathit{Expr}$
\item \textbf{undefine-define}\\$\mathrm{p}\left(\text{`'},\mathit{Ops},\varepsilon\right)$
\item \textbf{eliminate-introduce}\\$\mathrm{p}\left(\text{`'},\mathit{ObjectFactory},\varepsilon\right)$
\item \textbf{eliminate-introduce}\\$\mathrm{p}\left(\text{`'},\mathit{package-info},\varphi\right)$
\item \textbf{unchain-chain}\\$\mathrm{p}\left(\text{`'},\mathit{Expr},\mathit{Apply}\right)$
\item \textbf{unchain-chain}\\$\mathrm{p}\left(\text{`'},\mathit{Expr},\mathit{Argument}\right)$
\item \textbf{unchain-chain}\\$\mathrm{p}\left(\text{`'},\mathit{Expr},\mathit{Binary}\right)$
\item \textbf{unchain-chain}\\$\mathrm{p}\left(\text{`'},\mathit{Expr},\mathit{IfThenElse}\right)$
\item \textbf{unchain-chain}\\$\mathrm{p}\left(\text{`'},\mathit{Expr},\mathit{Literal}\right)$
\item \textbf{unlabel-designate}\\$\mathrm{p}\left(\fbox{\text{`Apply'}},\mathit{Expr},\mathrm{seq}\left(\left[str, \star \left(\mathit{Expr}\right)\right]\right)\right)$
\item \textbf{unlabel-designate}\\$\mathrm{p}\left(\fbox{\text{`Argument'}},\mathit{Expr},str\right)$
\item \textbf{unlabel-designate}\\$\mathrm{p}\left(\fbox{\text{`Binary'}},\mathit{Expr},\mathrm{seq}\left(\left[\mathit{Ops}, \mathit{Expr}, \mathit{Expr}\right]\right)\right)$
\item \textbf{unlabel-designate}\\$\mathrm{p}\left(\fbox{\text{`IfThenElse'}},\mathit{Expr},\mathrm{seq}\left(\left[\mathit{Expr}, \mathit{Expr}, \mathit{Expr}\right]\right)\right)$
\item \textbf{unlabel-designate}\\$\mathrm{p}\left(\fbox{\text{`Literal'}},\mathit{Expr},int\right)$
\item \textbf{extract-inline}  in $\mathit{Expr}$\\$\mathrm{p}\left(\text{`'},\mathit{Expr_1},\mathrm{seq}\left(\left[str, \star \left(\mathit{Expr}\right)\right]\right)\right)$
\item \textbf{extract-inline}  in $\mathit{Expr}$\\$\mathrm{p}\left(\text{`'},\mathit{Expr_2},\mathrm{seq}\left(\left[\mathit{Ops}, \mathit{Expr}, \mathit{Expr}\right]\right)\right)$
\item \textbf{extract-inline}  in $\mathit{Expr}$\\$\mathrm{p}\left(\text{`'},\mathit{Expr_3},\mathrm{seq}\left(\left[\mathit{Expr}, \mathit{Expr}, \mathit{Expr}\right]\right)\right)$
\end{itemize}}

\section{Grammar in ANF}

\footnotesize\begin{center}\begin{tabular}{|l|c|}\hline
\multicolumn{1}{|>{\columncolor[gray]{.9}}c|}{\footnotesize \textbf{Production rule}} &
\multicolumn{1}{>{\columncolor[gray]{.9}}c|}{\footnotesize \textbf{Production signature}}
\\\hline
$\mathrm{p}\left(\text{`'},\mathit{Expr},\mathit{Expr_1}\right)$	&	$\{ \langle \mathit{Expr_1}, 1\rangle\}$\\
$\mathrm{p}\left(\text{`'},\mathit{Expr},str\right)$	&	$\{ \langle str, 1\rangle\}$\\
$\mathrm{p}\left(\text{`'},\mathit{Expr},\mathit{Expr_2}\right)$	&	$\{ \langle \mathit{Expr_2}, 1\rangle\}$\\
$\mathrm{p}\left(\text{`'},\mathit{Expr},\mathit{Expr_3}\right)$	&	$\{ \langle \mathit{Expr_3}, 1\rangle\}$\\
$\mathrm{p}\left(\text{`'},\mathit{Expr},int\right)$	&	$\{ \langle int, 1\rangle\}$\\
$\mathrm{p}\left(\text{`'},\mathit{Function},\mathrm{seq}\left(\left[str, \star \left(str\right), \mathit{Expr}\right]\right)\right)$	&	$\{ \langle \mathit{Expr}, 1\rangle, \langle str, 1{*}\rangle\}$\\
$\mathrm{p}\left(\text{`'},\mathit{Program},\star \left(\mathit{Function}\right)\right)$	&	$\{ \langle \mathit{Function}, {*}\rangle\}$\\
$\mathrm{p}\left(\text{`'},\mathit{Expr_1},\mathrm{seq}\left(\left[str, \star \left(\mathit{Expr}\right)\right]\right)\right)$	&	$\{ \langle str, 1\rangle, \langle \mathit{Expr}, {*}\rangle\}$\\
$\mathrm{p}\left(\text{`'},\mathit{Expr_2},\mathrm{seq}\left(\left[\mathit{Ops}, \mathit{Expr}, \mathit{Expr}\right]\right)\right)$	&	$\{ \langle \mathit{Ops}, 1\rangle, \langle \mathit{Expr}, 11\rangle\}$\\
$\mathrm{p}\left(\text{`'},\mathit{Expr_3},\mathrm{seq}\left(\left[\mathit{Expr}, \mathit{Expr}, \mathit{Expr}\right]\right)\right)$	&	$\{ \langle \mathit{Expr}, 111\rangle\}$\\
\hline\end{tabular}\end{center}

\section{Nominal resolution}

Production rules are matched as follows (ANF on the left, master grammar on the right):
\begin{eqnarray*}
\mathrm{p}\left(\text{`'},\mathit{Expr},\mathit{Expr_1}\right) & \bumpeq & \mathrm{p}\left(\text{`'},\mathit{expression},\mathit{apply}\right) \\
\mathrm{p}\left(\text{`'},\mathit{Expr},str\right) & \bumpeq & \mathrm{p}\left(\text{`'},\mathit{expression},str\right) \\
\mathrm{p}\left(\text{`'},\mathit{Expr},\mathit{Expr_2}\right) & \bumpeq & \mathrm{p}\left(\text{`'},\mathit{expression},\mathit{binary}\right) \\
\mathrm{p}\left(\text{`'},\mathit{Expr},\mathit{Expr_3}\right) & \bumpeq & \mathrm{p}\left(\text{`'},\mathit{expression},\mathit{conditional}\right) \\
\mathrm{p}\left(\text{`'},\mathit{Expr},int\right) & \bumpeq & \mathrm{p}\left(\text{`'},\mathit{expression},int\right) \\
\mathrm{p}\left(\text{`'},\mathit{Function},\mathrm{seq}\left(\left[str, \star \left(str\right), \mathit{Expr}\right]\right)\right) & \Bumpeq & \mathrm{p}\left(\text{`'},\mathit{function},\mathrm{seq}\left(\left[str, \plus \left(str\right), \mathit{expression}\right]\right)\right) \\
\mathrm{p}\left(\text{`'},\mathit{Program},\star \left(\mathit{Function}\right)\right) & \Bumpeq & \mathrm{p}\left(\text{`'},\mathit{program},\plus \left(\mathit{function}\right)\right) \\
\mathrm{p}\left(\text{`'},\mathit{Expr_1},\mathrm{seq}\left(\left[str, \star \left(\mathit{Expr}\right)\right]\right)\right) & \Bumpeq & \mathrm{p}\left(\text{`'},\mathit{apply},\mathrm{seq}\left(\left[str, \plus \left(\mathit{expression}\right)\right]\right)\right) \\
\mathrm{p}\left(\text{`'},\mathit{Expr_2},\mathrm{seq}\left(\left[\mathit{Ops}, \mathit{Expr}, \mathit{Expr}\right]\right)\right) & \Bumpeq & \mathrm{p}\left(\text{`'},\mathit{binary},\mathrm{seq}\left(\left[\mathit{expression}, \mathit{operator}, \mathit{expression}\right]\right)\right) \\
\mathrm{p}\left(\text{`'},\mathit{Expr_3},\mathrm{seq}\left(\left[\mathit{Expr}, \mathit{Expr}, \mathit{Expr}\right]\right)\right) & \bumpeq & \mathrm{p}\left(\text{`'},\mathit{conditional},\mathrm{seq}\left(\left[\mathit{expression}, \mathit{expression}, \mathit{expression}\right]\right)\right) \\
\end{eqnarray*}
This yields the following nominal mapping:
\begin{align*}\mathit{jaxb} \:\diamond\: \mathit{master} =\:& \{\langle \mathit{Expr_2},\mathit{binary}\rangle,\\
 & \langle \mathit{Expr_3},\mathit{conditional}\rangle,\\
 & \langle int,int\rangle,\\
 & \langle \mathit{Function},\mathit{function}\rangle,\\
 & \langle str,str\rangle,\\
 & \langle \mathit{Program},\mathit{program}\rangle,\\
 & \langle \mathit{Expr},\mathit{expression}\rangle,\\
 & \langle \mathit{Expr_1},\mathit{apply}\rangle,\\
 & \langle \mathit{Ops},\mathit{operator}\rangle\}\end{align*}
 Which is exercised with these grammar transformation steps:

{\footnotesize\begin{itemize}
\item \textbf{renameN-renameN} $\mathit{Expr_2}$ to $\mathit{binary}$
\item \textbf{renameN-renameN} $\mathit{Expr_3}$ to $\mathit{conditional}$
\item \textbf{renameN-renameN} $\mathit{Function}$ to $\mathit{function}$
\item \textbf{renameN-renameN} $\mathit{Program}$ to $\mathit{program}$
\item \textbf{renameN-renameN} $\mathit{Expr}$ to $\mathit{expression}$
\item \textbf{renameN-renameN} $\mathit{Expr_1}$ to $\mathit{apply}$
\item \textbf{renameN-renameN} $\mathit{Ops}$ to $\mathit{operator}$
\end{itemize}}

\section{Structural resolution}
{\footnotesize\begin{itemize}
\item \textbf{narrow-widen}  in $\mathit{function}$\\$\star \left(str\right)$\\$\plus \left(str\right)$
\item \textbf{narrow-widen}  in $\mathit{program}$\\$\star \left(\mathit{function}\right)$\\$\plus \left(\mathit{function}\right)$
\item \textbf{narrow-widen}  in $\mathit{apply}$\\$\star \left(\mathit{expression}\right)$\\$\plus \left(\mathit{expression}\right)$
\item \textbf{permute-permute}\\$\mathrm{p}\left(\text{`'},\mathit{binary},\mathrm{seq}\left(\left[\mathit{operator}, \mathit{expression}, \mathit{expression}\right]\right)\right)$\\$\mathrm{p}\left(\text{`'},\mathit{binary},\mathrm{seq}\left(\left[\mathit{expression}, \mathit{operator}, \mathit{expression}\right]\right)\right)$
\end{itemize}}

\chapter{Java Object Model}

 Source name: \textbf{om}

\section{Source grammar}

\begin{itemize}\item Source artifact: \href{http://github.com/grammarware/slps/blob/master/topics/fl/java1/types/Apply.java}{topics/fl/java1/types/Apply.java}\item Source artifact: \href{http://github.com/grammarware/slps/blob/master/topics/fl/java1/types/Argument.java}{topics/fl/java1/types/Argument.java}\item Source artifact: \href{http://github.com/grammarware/slps/blob/master/topics/fl/java1/types/Binary.java}{topics/fl/java1/types/Binary.java}\item Source artifact: \href{http://github.com/grammarware/slps/blob/master/topics/fl/java1/types/Expr.java}{topics/fl/java1/types/Expr.java}\item Source artifact: \href{http://github.com/grammarware/slps/blob/master/topics/fl/java1/types/Function.java}{topics/fl/java1/types/Function.java}\item Source artifact: \href{http://github.com/grammarware/slps/blob/master/topics/fl/java1/types/IfThenElse.java}{topics/fl/java1/types/IfThenElse.java}\item Source artifact: \href{http://github.com/grammarware/slps/blob/master/topics/fl/java1/types/Literal.java}{topics/fl/java1/types/Literal.java}\item Source artifact: \href{http://github.com/grammarware/slps/blob/master/topics/fl/java1/types/Ops.java}{topics/fl/java1/types/Ops.java}\item Source artifact: \href{http://github.com/grammarware/slps/blob/master/topics/fl/java1/types/Program.java}{topics/fl/java1/types/Program.java}\item Source artifact: \href{http://github.com/grammarware/slps/blob/master/topics/fl/java1/types/Visitor.java}{topics/fl/java1/types/Visitor.java}\item Grammar extractor: \href{http://github.com/grammarware/slps/blob/master/topics/extraction/java2bgf/slps/java2bgf/Tool.java}{topics/extraction/java2bgf/slps/java2bgf/Tool.java}\end{itemize}

\footnotesize\begin{center}\begin{tabular}{|l|}\hline
\multicolumn{1}{|>{\columncolor[gray]{.9}}c|}{\footnotesize \textbf{Production rules}}
\\\hline
$\mathrm{p}(\text{`'},\mathit{Apply},\mathrm{seq}\left(\left[\mathrm{sel}\left(\text{`name'},str\right), \mathrm{sel}\left(\text{`args'},\star \left(\mathit{Expr}\right)\right)\right]\right))$	\\
$\mathrm{p}(\text{`'},\mathit{Argument},\mathrm{sel}\left(\text{`name'},str\right))$	\\
$\mathrm{p}(\text{`'},\mathit{Binary},\mathrm{seq}\left(\left[\mathrm{sel}\left(\text{`ops'},\mathit{Ops}\right), \mathrm{sel}\left(\text{`left'},\mathit{Expr}\right), \mathrm{sel}\left(\text{`right'},\mathit{Expr}\right)\right]\right))$	\\
$\mathrm{p}(\text{`'},\mathit{Expr},\mathrm{choice}([\mathit{Apply},$\\$\qquad\qquad\mathit{Argument},$\\$\qquad\qquad\mathit{Binary},$\\$\qquad\qquad\mathit{IfThenElse},$\\$\qquad\qquad\mathit{Literal}]))$	\\
$\mathrm{p}(\text{`'},\mathit{Function},\mathrm{seq}\left(\left[\mathrm{sel}\left(\text{`name'},str\right), \mathrm{sel}\left(\text{`args'},\star \left(str\right)\right), \mathrm{sel}\left(\text{`rhs'},\mathit{Expr}\right)\right]\right))$	\\
$\mathrm{p}(\text{`'},\mathit{IfThenElse},\mathrm{seq}\left(\left[\mathrm{sel}\left(\text{`ifExpr'},\mathit{Expr}\right), \mathrm{sel}\left(\text{`thenExpr'},\mathit{Expr}\right), \mathrm{sel}\left(\text{`elseExpr'},\mathit{Expr}\right)\right]\right))$	\\
$\mathrm{p}(\text{`'},\mathit{Literal},\mathrm{sel}\left(\text{`info'},int\right))$	\\
$\mathrm{p}(\text{`'},\mathit{Ops},\mathrm{choice}([\mathrm{sel}\left(\text{`Equal'},\varepsilon\right),$\\$\qquad\qquad\mathrm{sel}\left(\text{`Plus'},\varepsilon\right),$\\$\qquad\qquad\mathrm{sel}\left(\text{`Minus'},\varepsilon\right)]))$	\\
$\mathrm{p}(\text{`'},\mathit{Program},\mathrm{sel}\left(\text{`functions'},\star \left(\mathit{Function}\right)\right))$	\\
$\mathrm{p}(\text{`'},\mathit{Visitor},\varphi)$	\\
\hline\end{tabular}\end{center}

\section{Normalizations}
{\footnotesize\begin{itemize}
\item \textbf{reroot-reroot} $\left[\right]$ to $\left[\mathit{Program}\right]$
\item \textbf{unlabel-designate}\\$\mathrm{p}\left(\fbox{\text{`name'}},\mathit{Argument},str\right)$
\item \textbf{unlabel-designate}\\$\mathrm{p}\left(\fbox{\text{`info'}},\mathit{Literal},int\right)$
\item \textbf{unlabel-designate}\\$\mathrm{p}\left(\fbox{\text{`functions'}},\mathit{Program},\star \left(\mathit{Function}\right)\right)$
\item \textbf{anonymize-deanonymize}\\$\mathrm{p}\left(\text{`'},\mathit{IfThenElse},\mathrm{seq}\left(\left[\fbox{$\mathrm{sel}\left(\text{`ifExpr'},\mathit{Expr}\right)$}, \fbox{$\mathrm{sel}\left(\text{`thenExpr'},\mathit{Expr}\right)$}, \fbox{$\mathrm{sel}\left(\text{`elseExpr'},\mathit{Expr}\right)$}\right]\right)\right)$
\item \textbf{anonymize-deanonymize}\\$\mathrm{p}\left(\text{`'},\mathit{Ops},\mathrm{choice}\left(\left[\fbox{$\mathrm{sel}\left(\text{`Equal'},\varepsilon\right)$}, \fbox{$\mathrm{sel}\left(\text{`Plus'},\varepsilon\right)$}, \fbox{$\mathrm{sel}\left(\text{`Minus'},\varepsilon\right)$}\right]\right)\right)$
\item \textbf{anonymize-deanonymize}\\$\mathrm{p}\left(\text{`'},\mathit{Apply},\mathrm{seq}\left(\left[\fbox{$\mathrm{sel}\left(\text{`name'},str\right)$}, \fbox{$\mathrm{sel}\left(\text{`args'},\star \left(\mathit{Expr}\right)\right)$}\right]\right)\right)$
\item \textbf{anonymize-deanonymize}\\$\mathrm{p}\left(\text{`'},\mathit{Function},\mathrm{seq}\left(\left[\fbox{$\mathrm{sel}\left(\text{`name'},str\right)$}, \fbox{$\mathrm{sel}\left(\text{`args'},\star \left(str\right)\right)$}, \fbox{$\mathrm{sel}\left(\text{`rhs'},\mathit{Expr}\right)$}\right]\right)\right)$
\item \textbf{anonymize-deanonymize}\\$\mathrm{p}\left(\text{`'},\mathit{Binary},\mathrm{seq}\left(\left[\fbox{$\mathrm{sel}\left(\text{`ops'},\mathit{Ops}\right)$}, \fbox{$\mathrm{sel}\left(\text{`left'},\mathit{Expr}\right)$}, \fbox{$\mathrm{sel}\left(\text{`right'},\mathit{Expr}\right)$}\right]\right)\right)$
\item \textbf{vertical-horizontal}  in $\mathit{Expr}$
\item \textbf{eliminate-introduce}\\$\mathrm{p}\left(\text{`'},\mathit{Visitor},\varphi\right)$
\item \textbf{undefine-define}\\$\mathrm{p}\left(\text{`'},\mathit{Ops},\varepsilon\right)$
\item \textbf{unchain-chain}\\$\mathrm{p}\left(\text{`'},\mathit{Expr},\mathit{Apply}\right)$
\item \textbf{unchain-chain}\\$\mathrm{p}\left(\text{`'},\mathit{Expr},\mathit{Argument}\right)$
\item \textbf{unchain-chain}\\$\mathrm{p}\left(\text{`'},\mathit{Expr},\mathit{Binary}\right)$
\item \textbf{unchain-chain}\\$\mathrm{p}\left(\text{`'},\mathit{Expr},\mathit{IfThenElse}\right)$
\item \textbf{unchain-chain}\\$\mathrm{p}\left(\text{`'},\mathit{Expr},\mathit{Literal}\right)$
\item \textbf{unlabel-designate}\\$\mathrm{p}\left(\fbox{\text{`Apply'}},\mathit{Expr},\mathrm{seq}\left(\left[str, \star \left(\mathit{Expr}\right)\right]\right)\right)$
\item \textbf{unlabel-designate}\\$\mathrm{p}\left(\fbox{\text{`Argument'}},\mathit{Expr},str\right)$
\item \textbf{unlabel-designate}\\$\mathrm{p}\left(\fbox{\text{`Binary'}},\mathit{Expr},\mathrm{seq}\left(\left[\mathit{Ops}, \mathit{Expr}, \mathit{Expr}\right]\right)\right)$
\item \textbf{unlabel-designate}\\$\mathrm{p}\left(\fbox{\text{`IfThenElse'}},\mathit{Expr},\mathrm{seq}\left(\left[\mathit{Expr}, \mathit{Expr}, \mathit{Expr}\right]\right)\right)$
\item \textbf{unlabel-designate}\\$\mathrm{p}\left(\fbox{\text{`Literal'}},\mathit{Expr},int\right)$
\item \textbf{extract-inline}  in $\mathit{Expr}$\\$\mathrm{p}\left(\text{`'},\mathit{Expr_1},\mathrm{seq}\left(\left[str, \star \left(\mathit{Expr}\right)\right]\right)\right)$
\item \textbf{extract-inline}  in $\mathit{Expr}$\\$\mathrm{p}\left(\text{`'},\mathit{Expr_2},\mathrm{seq}\left(\left[\mathit{Ops}, \mathit{Expr}, \mathit{Expr}\right]\right)\right)$
\item \textbf{extract-inline}  in $\mathit{Expr}$\\$\mathrm{p}\left(\text{`'},\mathit{Expr_3},\mathrm{seq}\left(\left[\mathit{Expr}, \mathit{Expr}, \mathit{Expr}\right]\right)\right)$
\end{itemize}}

\section{Grammar in ANF}

\footnotesize\begin{center}\begin{tabular}{|l|c|}\hline
\multicolumn{1}{|>{\columncolor[gray]{.9}}c|}{\footnotesize \textbf{Production rule}} &
\multicolumn{1}{>{\columncolor[gray]{.9}}c|}{\footnotesize \textbf{Production signature}}
\\\hline
$\mathrm{p}\left(\text{`'},\mathit{Expr},\mathit{Expr_1}\right)$	&	$\{ \langle \mathit{Expr_1}, 1\rangle\}$\\
$\mathrm{p}\left(\text{`'},\mathit{Expr},str\right)$	&	$\{ \langle str, 1\rangle\}$\\
$\mathrm{p}\left(\text{`'},\mathit{Expr},\mathit{Expr_2}\right)$	&	$\{ \langle \mathit{Expr_2}, 1\rangle\}$\\
$\mathrm{p}\left(\text{`'},\mathit{Expr},\mathit{Expr_3}\right)$	&	$\{ \langle \mathit{Expr_3}, 1\rangle\}$\\
$\mathrm{p}\left(\text{`'},\mathit{Expr},int\right)$	&	$\{ \langle int, 1\rangle\}$\\
$\mathrm{p}\left(\text{`'},\mathit{Function},\mathrm{seq}\left(\left[str, \star \left(str\right), \mathit{Expr}\right]\right)\right)$	&	$\{ \langle \mathit{Expr}, 1\rangle, \langle str, 1{*}\rangle\}$\\
$\mathrm{p}\left(\text{`'},\mathit{Program},\star \left(\mathit{Function}\right)\right)$	&	$\{ \langle \mathit{Function}, {*}\rangle\}$\\
$\mathrm{p}\left(\text{`'},\mathit{Expr_1},\mathrm{seq}\left(\left[str, \star \left(\mathit{Expr}\right)\right]\right)\right)$	&	$\{ \langle str, 1\rangle, \langle \mathit{Expr}, {*}\rangle\}$\\
$\mathrm{p}\left(\text{`'},\mathit{Expr_2},\mathrm{seq}\left(\left[\mathit{Ops}, \mathit{Expr}, \mathit{Expr}\right]\right)\right)$	&	$\{ \langle \mathit{Ops}, 1\rangle, \langle \mathit{Expr}, 11\rangle\}$\\
$\mathrm{p}\left(\text{`'},\mathit{Expr_3},\mathrm{seq}\left(\left[\mathit{Expr}, \mathit{Expr}, \mathit{Expr}\right]\right)\right)$	&	$\{ \langle \mathit{Expr}, 111\rangle\}$\\
\hline\end{tabular}\end{center}

\section{Nominal resolution}

Production rules are matched as follows (ANF on the left, master grammar on the right):
\begin{eqnarray*}
\mathrm{p}\left(\text{`'},\mathit{Expr},\mathit{Expr_1}\right) & \bumpeq & \mathrm{p}\left(\text{`'},\mathit{expression},\mathit{apply}\right) \\
\mathrm{p}\left(\text{`'},\mathit{Expr},str\right) & \bumpeq & \mathrm{p}\left(\text{`'},\mathit{expression},str\right) \\
\mathrm{p}\left(\text{`'},\mathit{Expr},\mathit{Expr_2}\right) & \bumpeq & \mathrm{p}\left(\text{`'},\mathit{expression},\mathit{binary}\right) \\
\mathrm{p}\left(\text{`'},\mathit{Expr},\mathit{Expr_3}\right) & \bumpeq & \mathrm{p}\left(\text{`'},\mathit{expression},\mathit{conditional}\right) \\
\mathrm{p}\left(\text{`'},\mathit{Expr},int\right) & \bumpeq & \mathrm{p}\left(\text{`'},\mathit{expression},int\right) \\
\mathrm{p}\left(\text{`'},\mathit{Function},\mathrm{seq}\left(\left[str, \star \left(str\right), \mathit{Expr}\right]\right)\right) & \Bumpeq & \mathrm{p}\left(\text{`'},\mathit{function},\mathrm{seq}\left(\left[str, \plus \left(str\right), \mathit{expression}\right]\right)\right) \\
\mathrm{p}\left(\text{`'},\mathit{Program},\star \left(\mathit{Function}\right)\right) & \Bumpeq & \mathrm{p}\left(\text{`'},\mathit{program},\plus \left(\mathit{function}\right)\right) \\
\mathrm{p}\left(\text{`'},\mathit{Expr_1},\mathrm{seq}\left(\left[str, \star \left(\mathit{Expr}\right)\right]\right)\right) & \Bumpeq & \mathrm{p}\left(\text{`'},\mathit{apply},\mathrm{seq}\left(\left[str, \plus \left(\mathit{expression}\right)\right]\right)\right) \\
\mathrm{p}\left(\text{`'},\mathit{Expr_2},\mathrm{seq}\left(\left[\mathit{Ops}, \mathit{Expr}, \mathit{Expr}\right]\right)\right) & \Bumpeq & \mathrm{p}\left(\text{`'},\mathit{binary},\mathrm{seq}\left(\left[\mathit{expression}, \mathit{operator}, \mathit{expression}\right]\right)\right) \\
\mathrm{p}\left(\text{`'},\mathit{Expr_3},\mathrm{seq}\left(\left[\mathit{Expr}, \mathit{Expr}, \mathit{Expr}\right]\right)\right) & \bumpeq & \mathrm{p}\left(\text{`'},\mathit{conditional},\mathrm{seq}\left(\left[\mathit{expression}, \mathit{expression}, \mathit{expression}\right]\right)\right) \\
\end{eqnarray*}
This yields the following nominal mapping:
\begin{align*}\mathit{om} \:\diamond\: \mathit{master} =\:& \{\langle \mathit{Expr_2},\mathit{binary}\rangle,\\
 & \langle \mathit{Expr_3},\mathit{conditional}\rangle,\\
 & \langle int,int\rangle,\\
 & \langle \mathit{Function},\mathit{function}\rangle,\\
 & \langle str,str\rangle,\\
 & \langle \mathit{Program},\mathit{program}\rangle,\\
 & \langle \mathit{Expr},\mathit{expression}\rangle,\\
 & \langle \mathit{Expr_1},\mathit{apply}\rangle,\\
 & \langle \mathit{Ops},\mathit{operator}\rangle\}\end{align*}
 Which is exercised with these grammar transformation steps:

{\footnotesize\begin{itemize}
\item \textbf{renameN-renameN} $\mathit{Expr_2}$ to $\mathit{binary}$
\item \textbf{renameN-renameN} $\mathit{Expr_3}$ to $\mathit{conditional}$
\item \textbf{renameN-renameN} $\mathit{Function}$ to $\mathit{function}$
\item \textbf{renameN-renameN} $\mathit{Program}$ to $\mathit{program}$
\item \textbf{renameN-renameN} $\mathit{Expr}$ to $\mathit{expression}$
\item \textbf{renameN-renameN} $\mathit{Expr_1}$ to $\mathit{apply}$
\item \textbf{renameN-renameN} $\mathit{Ops}$ to $\mathit{operator}$
\end{itemize}}

\section{Structural resolution}
{\footnotesize\begin{itemize}
\item \textbf{narrow-widen}  in $\mathit{function}$\\$\star \left(str\right)$\\$\plus \left(str\right)$
\item \textbf{narrow-widen}  in $\mathit{program}$\\$\star \left(\mathit{function}\right)$\\$\plus \left(\mathit{function}\right)$
\item \textbf{narrow-widen}  in $\mathit{apply}$\\$\star \left(\mathit{expression}\right)$\\$\plus \left(\mathit{expression}\right)$
\item \textbf{permute-permute}\\$\mathrm{p}\left(\text{`'},\mathit{binary},\mathrm{seq}\left(\left[\mathit{operator}, \mathit{expression}, \mathit{expression}\right]\right)\right)$\\$\mathrm{p}\left(\text{`'},\mathit{binary},\mathrm{seq}\left(\left[\mathit{expression}, \mathit{operator}, \mathit{expression}\right]\right)\right)$
\end{itemize}}

\chapter{PyParsing in Python}

 Source name: \textbf{python}

\section{Source grammar}

\begin{itemize}\item Source artifact: \href{http://github.com/grammarware/slps/blob/master/topics/fl/python/parser.py}{topics/fl/python/parser.py}\item Grammar extractor: \href{http://github.com/grammarware/slps/blob/master/shared/rascal/src/extract/Python2BGF.rsc}{shared/rascal/src/extract/Python2BGF.rsc}\end{itemize}

\footnotesize\begin{center}\begin{tabular}{|l|}\hline
\multicolumn{1}{|>{\columncolor[gray]{.9}}c|}{\footnotesize \textbf{Production rules}}
\\\hline
$\mathrm{p}(\text{`'},\mathit{\_ Literal},\mathit{Literal})$	\\
$\mathrm{p}(\text{`'},\mathit{\_ IF},\text{`if'})$	\\
$\mathrm{p}(\text{`'},\mathit{\_ THEN},\text{`then'})$	\\
$\mathrm{p}(\text{`'},\mathit{\_ ELSE},\text{`else'})$	\\
$\mathrm{p}(\text{`'},\mathit{name},str)$	\\
$\mathrm{p}(\text{`'},\mathit{literal},\mathrm{seq}\left(\left[\opt \left(\text{`-'}\right), int\right]\right))$	\\
$\mathrm{p}(\text{`'},\mathit{atom},\mathrm{choice}([\mathit{name},$\\$\qquad\qquad\mathit{literal},$\\$\qquad\qquad\mathrm{seq}\left(\left[\text{`('}, \mathit{expr}, \text{`)'}\right]\right)]))$	\\
$\mathrm{p}(\text{`'},\mathit{ifThenElse},\mathrm{seq}\left(\left[\mathit{\_ IF}, \mathit{expr}, \mathit{\_ THEN}, \mathit{expr}, \mathit{\_ ELSE}, \mathit{expr}\right]\right))$	\\
$\mathrm{p}(\text{`'},\mathit{operators},\mathrm{choice}([\text{`=='},$\\$\qquad\qquad\text{`+'},$\\$\qquad\qquad\text{`-'}]))$	\\
$\mathrm{p}(\text{`'},\mathit{binary},\mathrm{seq}\left(\left[\mathit{atom}, \star \left(\mathrm{seq}\left(\left[\mathit{operators}, \mathit{atom}\right]\right)\right)\right]\right))$	\\
$\mathrm{p}(\text{`'},\mathit{apply},\mathrm{seq}\left(\left[\mathit{name}, \plus \left(\mathit{atom}\right)\right]\right))$	\\
$\mathrm{p}(\text{`'},\mathit{expr},\mathrm{choice}([\mathit{binary},$\\$\qquad\qquad\mathit{apply},$\\$\qquad\qquad\mathit{ifThenElse}]))$	\\
$\mathrm{p}(\text{`'},\mathit{function},\mathrm{seq}\left(\left[\mathit{name}, \plus \left(\mathit{name}\right), \text{`='}, \mathit{expr}\right]\right))$	\\
$\mathrm{p}(\text{`'},\mathit{program},\mathrm{seq}\left(\left[\plus \left(\mathit{function}\right), \mathit{StringEnd}\right]\right))$	\\
\hline\end{tabular}\end{center}

\section{Mutations}
{\footnotesize\begin{itemize}
\item \textbf{unite-splitN} $expr$ \\$\mathrm{p}\left(\text{`'},\mathit{atom},\mathrm{choice}\left(\left[\mathit{name}, \mathit{literal}, \mathrm{seq}\left(\left[\text{`('}, \mathit{expr}, \text{`)'}\right]\right)\right]\right)\right)$
\item \textbf{designate-unlabel}\\$\mathrm{p}\left(\fbox{\text{`tmplabel'}},\mathit{binary},\mathrm{seq}\left(\left[\mathit{expr}, \star \left(\mathrm{seq}\left(\left[\mathit{operators}, \mathit{expr}\right]\right)\right)\right]\right)\right)$
\item \textbf{assoc-iterate}\\$\mathrm{p}\left(\text{`tmplabel'},\mathit{binary},\mathrm{seq}\left(\left[\mathit{expr}, \mathit{operators}, \mathit{expr}\right]\right)\right)$
\item \textbf{unlabel-designate}\\$\mathrm{p}\left(\fbox{\text{`tmplabel'}},\mathit{binary},\mathrm{seq}\left(\left[\mathit{expr}, \mathit{operators}, \mathit{expr}\right]\right)\right)$
\end{itemize}}

\section{Normalizations}
{\footnotesize\begin{itemize}
\item \textbf{reroot-reroot} $\left[\right]$ to $\left[\mathit{program}\right]$
\item \textbf{abstractize-concretize}\\$\mathrm{p}\left(\text{`'},\mathit{literal},\mathrm{seq}\left(\left[\opt \left(\fbox{$\text{`-'}$}\right), int\right]\right)\right)$
\item \textbf{abstractize-concretize}\\$\mathrm{p}\left(\text{`'},\mathit{operators},\mathrm{choice}\left(\left[\fbox{$\text{`=='}$}, \fbox{$\text{`+'}$}, \fbox{$\text{`-'}$}\right]\right)\right)$
\item \textbf{abstractize-concretize}\\$\mathrm{p}\left(\text{`'},\mathit{\_ IF},\fbox{$\text{`if'}$}\right)$
\item \textbf{abstractize-concretize}\\$\mathrm{p}\left(\text{`'},\mathit{expr},\mathrm{choice}\left(\left[\mathit{name}, \mathit{literal}, \mathrm{seq}\left(\left[\fbox{$\text{`('}$}, \mathit{expr}, \fbox{$\text{`)'}$}\right]\right)\right]\right)\right)$
\item \textbf{abstractize-concretize}\\$\mathrm{p}\left(\text{`'},\mathit{\_ ELSE},\fbox{$\text{`else'}$}\right)$
\item \textbf{abstractize-concretize}\\$\mathrm{p}\left(\text{`'},\mathit{\_ THEN},\fbox{$\text{`then'}$}\right)$
\item \textbf{abstractize-concretize}\\$\mathrm{p}\left(\text{`'},\mathit{function},\mathrm{seq}\left(\left[\mathit{name}, \plus \left(\mathit{name}\right), \fbox{$\text{`='}$}, \mathit{expr}\right]\right)\right)$
\item \textbf{vertical-horizontal}  in $\mathit{expr}$
\item \textbf{undefine-define}\\$\mathrm{p}\left(\text{`'},\mathit{\_ IF},\varepsilon\right)$
\item \textbf{undefine-define}\\$\mathrm{p}\left(\text{`'},\mathit{\_ THEN},\varepsilon\right)$
\item \textbf{undefine-define}\\$\mathrm{p}\left(\text{`'},\mathit{\_ ELSE},\varepsilon\right)$
\item \textbf{undefine-define}\\$\mathrm{p}\left(\text{`'},\mathit{operators},\varepsilon\right)$
\item \textbf{unchain-chain}\\$\mathrm{p}\left(\text{`'},\mathit{expr},\mathit{literal}\right)$
\item \textbf{abridge-detour}\\$\mathrm{p}\left(\text{`'},\mathit{expr},\mathit{expr}\right)$
\item \textbf{unchain-chain}\\$\mathrm{p}\left(\text{`'},\mathit{expr},\mathit{binary}\right)$
\item \textbf{unchain-chain}\\$\mathrm{p}\left(\text{`'},\mathit{expr},\mathit{apply}\right)$
\item \textbf{unchain-chain}\\$\mathrm{p}\left(\text{`'},\mathit{expr},\mathit{ifThenElse}\right)$
\item \textbf{inline-extract}\\$\mathrm{p}\left(\text{`'},\mathit{name},str\right)$
\item \textbf{unlabel-designate}\\$\mathrm{p}\left(\fbox{\text{`literal'}},\mathit{expr},int\right)$
\item \textbf{unlabel-designate}\\$\mathrm{p}\left(\fbox{\text{`ifThenElse'}},\mathit{expr},\mathrm{seq}\left(\left[\mathit{\_ IF}, \mathit{expr}, \mathit{\_ THEN}, \mathit{expr}, \mathit{\_ ELSE}, \mathit{expr}\right]\right)\right)$
\item \textbf{unlabel-designate}\\$\mathrm{p}\left(\fbox{\text{`binary'}},\mathit{expr},\mathrm{seq}\left(\left[\mathit{expr}, \mathit{operators}, \mathit{expr}\right]\right)\right)$
\item \textbf{unlabel-designate}\\$\mathrm{p}\left(\fbox{\text{`apply'}},\mathit{expr},\mathrm{seq}\left(\left[str, \plus \left(\mathit{expr}\right)\right]\right)\right)$
\item \textbf{extract-inline}  in $\mathit{expr}$\\$\mathrm{p}\left(\text{`'},\mathit{expr_1},\mathrm{seq}\left(\left[\mathit{\_ IF}, \mathit{expr}, \mathit{\_ THEN}, \mathit{expr}, \mathit{\_ ELSE}, \mathit{expr}\right]\right)\right)$
\item \textbf{extract-inline}  in $\mathit{expr}$\\$\mathrm{p}\left(\text{`'},\mathit{expr_2},\mathrm{seq}\left(\left[\mathit{expr}, \mathit{operators}, \mathit{expr}\right]\right)\right)$
\item \textbf{extract-inline}  in $\mathit{expr}$\\$\mathrm{p}\left(\text{`'},\mathit{expr_3},\mathrm{seq}\left(\left[str, \plus \left(\mathit{expr}\right)\right]\right)\right)$
\end{itemize}}

\section{Grammar in ANF}

\footnotesize\begin{center}\begin{tabular}{|l|c|}\hline
\multicolumn{1}{|>{\columncolor[gray]{.9}}c|}{\footnotesize \textbf{Production rule}} &
\multicolumn{1}{>{\columncolor[gray]{.9}}c|}{\footnotesize \textbf{Production signature}}
\\\hline
$\mathrm{p}\left(\text{`'},\mathit{\_ Literal},\mathit{Literal}\right)$	&	$\{ \langle \mathit{Literal}, 1\rangle\}$\\
$\mathrm{p}\left(\text{`'},\mathit{expr},int\right)$	&	$\{ \langle int, 1\rangle\}$\\
$\mathrm{p}\left(\text{`'},\mathit{expr},str\right)$	&	$\{ \langle str, 1\rangle\}$\\
$\mathrm{p}\left(\text{`'},\mathit{expr},\mathit{expr_1}\right)$	&	$\{ \langle \mathit{expr_1}, 1\rangle\}$\\
$\mathrm{p}\left(\text{`'},\mathit{expr},\mathit{expr_2}\right)$	&	$\{ \langle \mathit{expr_2}, 1\rangle\}$\\
$\mathrm{p}\left(\text{`'},\mathit{expr},\mathit{expr_3}\right)$	&	$\{ \langle \mathit{expr_3}, 1\rangle\}$\\
$\mathrm{p}\left(\text{`'},\mathit{function},\mathrm{seq}\left(\left[str, \plus \left(str\right), \mathit{expr}\right]\right)\right)$	&	$\{ \langle str, 1{+}\rangle, \langle \mathit{expr}, 1\rangle\}$\\
$\mathrm{p}\left(\text{`'},\mathit{program},\mathrm{seq}\left(\left[\plus \left(\mathit{function}\right), \mathit{StringEnd}\right]\right)\right)$	&	$\{ \langle \mathit{function}, {+}\rangle, \langle \mathit{StringEnd}, 1\rangle\}$\\
$\mathrm{p}\left(\text{`'},\mathit{expr_1},\mathrm{seq}\left(\left[\mathit{\_ IF}, \mathit{expr}, \mathit{\_ THEN}, \mathit{expr}, \mathit{\_ ELSE}, \mathit{expr}\right]\right)\right)$	&	$\{ \langle \mathit{\_ IF}, 1\rangle, \langle \mathit{\_ THEN}, 1\rangle, \langle \mathit{expr}, 111\rangle, \langle \mathit{\_ ELSE}, 1\rangle\}$\\
$\mathrm{p}\left(\text{`'},\mathit{expr_2},\mathrm{seq}\left(\left[\mathit{expr}, \mathit{operators}, \mathit{expr}\right]\right)\right)$	&	$\{ \langle \mathit{expr}, 11\rangle, \langle \mathit{operators}, 1\rangle\}$\\
$\mathrm{p}\left(\text{`'},\mathit{expr_3},\mathrm{seq}\left(\left[str, \plus \left(\mathit{expr}\right)\right]\right)\right)$	&	$\{ \langle str, 1\rangle, \langle \mathit{expr}, {+}\rangle\}$\\
\hline\end{tabular}\end{center}

\section{Nominal resolution}

Production rules are matched as follows (ANF on the left, master grammar on the right):
\begin{eqnarray*}
\mathrm{p}\left(\text{`'},\mathit{\_ Literal},\mathit{Literal}\right) &  & \varnothing \\
\mathrm{p}\left(\text{`'},\mathit{expr},int\right) & \bumpeq & \mathrm{p}\left(\text{`'},\mathit{expression},int\right) \\
\mathrm{p}\left(\text{`'},\mathit{expr},str\right) & \bumpeq & \mathrm{p}\left(\text{`'},\mathit{expression},str\right) \\
\mathrm{p}\left(\text{`'},\mathit{expr},\mathit{expr_1}\right) & \bumpeq & \mathrm{p}\left(\text{`'},\mathit{expression},\mathit{conditional}\right) \\
\mathrm{p}\left(\text{`'},\mathit{expr},\mathit{expr_2}\right) & \bumpeq & \mathrm{p}\left(\text{`'},\mathit{expression},\mathit{binary}\right) \\
\mathrm{p}\left(\text{`'},\mathit{expr},\mathit{expr_3}\right) & \bumpeq & \mathrm{p}\left(\text{`'},\mathit{expression},\mathit{apply}\right) \\
\mathrm{p}\left(\text{`'},\mathit{function},\mathrm{seq}\left(\left[str, \plus \left(str\right), \mathit{expr}\right]\right)\right) & \bumpeq & \mathrm{p}\left(\text{`'},\mathit{function},\mathrm{seq}\left(\left[str, \plus \left(str\right), \mathit{expression}\right]\right)\right) \\
\mathrm{p}\left(\text{`'},\mathit{program},\mathrm{seq}\left(\left[\plus \left(\mathit{function}\right), \mathit{StringEnd}\right]\right)\right) & \Bumpeq & \mathrm{p}\left(\text{`'},\mathit{program},\plus \left(\mathit{function}\right)\right) \\
\mathrm{p}\left(\text{`'},\mathit{expr_1},\mathrm{seq}\left(\left[\mathit{\_ IF}, \mathit{expr}, \mathit{\_ THEN}, \mathit{expr}, \mathit{\_ ELSE}, \mathit{expr}\right]\right)\right) & \Bumpeq & \mathrm{p}\left(\text{`'},\mathit{conditional},\mathrm{seq}\left(\left[\mathit{expression}, \mathit{expression}, \mathit{expression}\right]\right)\right) \\
\mathrm{p}\left(\text{`'},\mathit{expr_2},\mathrm{seq}\left(\left[\mathit{expr}, \mathit{operators}, \mathit{expr}\right]\right)\right) & \bumpeq & \mathrm{p}\left(\text{`'},\mathit{binary},\mathrm{seq}\left(\left[\mathit{expression}, \mathit{operator}, \mathit{expression}\right]\right)\right) \\
\mathrm{p}\left(\text{`'},\mathit{expr_3},\mathrm{seq}\left(\left[str, \plus \left(\mathit{expr}\right)\right]\right)\right) & \bumpeq & \mathrm{p}\left(\text{`'},\mathit{apply},\mathrm{seq}\left(\left[str, \plus \left(\mathit{expression}\right)\right]\right)\right) \\
\end{eqnarray*}
This yields the following nominal mapping:
\begin{align*}\mathit{python} \:\diamond\: \mathit{master} =\:& \{\langle \mathit{expr_2},\mathit{binary}\rangle,\\
 & \langle \mathit{program},\mathit{program}\rangle,\\
 & \langle \mathit{function},\mathit{function}\rangle,\\
 & \langle \mathit{expr_1},\mathit{conditional}\rangle,\\
 & \langle \mathit{expr},\mathit{expression}\rangle,\\
 & \langle str,str\rangle,\\
 & \langle int,int\rangle,\\
 & \langle \mathit{StringEnd},\omega\rangle,\\
 & \langle \mathit{\_ ELSE},\omega\rangle,\\
 & \langle \mathit{\_ IF},\omega\rangle,\\
 & \langle \mathit{expr_3},\mathit{apply}\rangle,\\
 & \langle \mathit{\_ THEN},\omega\rangle,\\
 & \langle \mathit{operators},\mathit{operator}\rangle\}\end{align*}
 Which is exercised with these grammar transformation steps:

{\footnotesize\begin{itemize}
\item \textbf{renameN-renameN} $\mathit{expr_2}$ to $\mathit{binary}$
\item \textbf{renameN-renameN} $\mathit{expr_1}$ to $\mathit{conditional}$
\item \textbf{renameN-renameN} $\mathit{expr}$ to $\mathit{expression}$
\item \textbf{renameN-renameN} $\mathit{expr_3}$ to $\mathit{apply}$
\item \textbf{renameN-renameN} $\mathit{operators}$ to $\mathit{operator}$
\end{itemize}}

\section{Structural resolution}
{\footnotesize\begin{itemize}
\item \textbf{project-inject}\\$\mathrm{p}\left(\text{`'},\mathit{program},\mathrm{seq}\left(\left[\plus \left(\mathit{function}\right), \fbox{$\mathit{StringEnd}$}\right]\right)\right)$
\item \textbf{project-inject}\\$\mathrm{p}\left(\text{`'},\mathit{conditional},\mathrm{seq}\left(\left[\mathit{\_ IF}, \mathit{expression}, \mathit{\_ THEN}, \mathit{expression}, \fbox{$\mathit{\_ ELSE}$}, \mathit{expression}\right]\right)\right)$
\item \textbf{project-inject}\\$\mathrm{p}\left(\text{`'},\mathit{conditional},\mathrm{seq}\left(\left[\fbox{$\mathit{\_ IF}$}, \mathit{expression}, \mathit{\_ THEN}, \mathit{expression}, \mathit{expression}\right]\right)\right)$
\item \textbf{project-inject}\\$\mathrm{p}\left(\text{`'},\mathit{conditional},\mathrm{seq}\left(\left[\mathit{expression}, \fbox{$\mathit{\_ THEN}$}, \mathit{expression}, \mathit{expression}\right]\right)\right)$
\item \textbf{eliminate-introduce}\\$\mathrm{p}\left(\text{`'},\mathit{\_ Literal},\mathit{Literal}\right)$
\end{itemize}}

\chapter{Rascal Algebraic Data Type}

 Source name: \textbf{rascal-a}

\section{Source grammar}

\begin{itemize}\item Source artifact: \href{http://github.com/grammarware/slps/blob/master/topics/fl/rascal/Abstract.rsc}{topics/fl/rascal/Abstract.rsc}\item Grammar extractor: \href{http://github.com/grammarware/slps/blob/master/shared/rascal/src/extract/RascalADT2BGF.rsc}{shared/rascal/src/extract/RascalADT2BGF.rsc}\end{itemize}

\footnotesize\begin{center}\begin{tabular}{|l|}\hline
\multicolumn{1}{|>{\columncolor[gray]{.9}}c|}{\footnotesize \textbf{Production rules}}
\\\hline
$\mathrm{p}(\text{`prg'},\mathit{FLPrg},\mathrm{sel}\left(\text{`fs'},\star \left(\mathit{FLFun}\right)\right))$	\\
$\mathrm{p}(\text{`fun'},\mathit{FLFun},\mathrm{seq}\left(\left[\mathrm{sel}\left(\text{`f'},str\right), \mathrm{sel}\left(\text{`args'},\star \left(str\right)\right), \mathrm{sel}\left(\text{`body'},\mathit{FLExpr}\right)\right]\right))$	\\
$\mathrm{p}(\text{`'},\mathit{FLExpr},\mathrm{choice}([\mathrm{sel}\left(\text{`binary'},\mathrm{seq}\left(\left[\mathrm{sel}\left(\text{`e1'},\mathit{FLExpr}\right), \mathrm{sel}\left(\text{`op'},\mathit{FLOp}\right), \mathrm{sel}\left(\text{`e2'},\mathit{FLExpr}\right)\right]\right)\right),$\\$\qquad\qquad\mathrm{sel}\left(\text{`apply'},\mathrm{seq}\left(\left[\mathrm{sel}\left(\text{`f'},str\right), \mathrm{sel}\left(\text{`vargs'},\star \left(\mathit{FLExpr}\right)\right)\right]\right)\right),$\\$\qquad\qquad\mathrm{sel}\left(\text{`ifThenElse'},\mathrm{seq}\left(\left[\mathrm{sel}\left(\text{`c'},\mathit{FLExpr}\right), \mathrm{sel}\left(\text{`t'},\mathit{FLExpr}\right), \mathrm{sel}\left(\text{`e'},\mathit{FLExpr}\right)\right]\right)\right),$\\$\qquad\qquad\mathrm{sel}\left(\text{`argument'},\mathrm{sel}\left(\text{`a'},str\right)\right),$\\$\qquad\qquad\mathrm{sel}\left(\text{`literal'},\mathrm{sel}\left(\text{`i'},int\right)\right)]))$	\\
$\mathrm{p}(\text{`'},\mathit{FLOp},\mathrm{choice}([\mathrm{sel}\left(\text{`minus'},\varepsilon\right),$\\$\qquad\qquad\mathrm{sel}\left(\text{`plus'},\varepsilon\right),$\\$\qquad\qquad\mathrm{sel}\left(\text{`equal'},\varepsilon\right)]))$	\\
\hline\end{tabular}\end{center}

\section{Normalizations}
{\footnotesize\begin{itemize}
\item \textbf{reroot-reroot} $\left[\right]$ to $\left[\mathit{FLPrg}\right]$
\item \textbf{unlabel-designate}\\$\mathrm{p}\left(\fbox{\text{`prg'}},\mathit{FLPrg},\mathrm{sel}\left(\text{`fs'},\star \left(\mathit{FLFun}\right)\right)\right)$
\item \textbf{unlabel-designate}\\$\mathrm{p}\left(\fbox{\text{`fun'}},\mathit{FLFun},\mathrm{seq}\left(\left[\mathrm{sel}\left(\text{`f'},str\right), \mathrm{sel}\left(\text{`args'},\star \left(str\right)\right), \mathrm{sel}\left(\text{`body'},\mathit{FLExpr}\right)\right]\right)\right)$
\item \textbf{anonymize-deanonymize}\\$\mathrm{p}\left(\text{`'},\mathit{FLOp},\mathrm{choice}\left(\left[\fbox{$\mathrm{sel}\left(\text{`minus'},\varepsilon\right)$}, \fbox{$\mathrm{sel}\left(\text{`plus'},\varepsilon\right)$}, \fbox{$\mathrm{sel}\left(\text{`equal'},\varepsilon\right)$}\right]\right)\right)$
\item \textbf{anonymize-deanonymize}\\$\mathrm{p}\left(\text{`'},\mathit{FLExpr},\mathrm{choice}\left(\left[\fbox{$\mathrm{sel}\left(\text{`binary'},\mathrm{seq}\left(\left[\fbox{$\mathrm{sel}\left(\text{`e1'},\mathit{FLExpr}\right)$}, \fbox{$\mathrm{sel}\left(\text{`op'},\mathit{FLOp}\right)$}, \fbox{$\mathrm{sel}\left(\text{`e2'},\mathit{FLExpr}\right)$}\right]\right)\right)$}, \fbox{$\mathrm{sel}\left(\text{`apply'},\mathrm{seq}\left(\left[\fbox{$\mathrm{sel}\left(\text{`f'},str\right)$}, \fbox{$\mathrm{sel}\left(\text{`vargs'},\star \left(\mathit{FLExpr}\right)\right)$}\right]\right)\right)$}, \fbox{$\mathrm{sel}\left(\text{`ifThenElse'},\mathrm{seq}\left(\left[\fbox{$\mathrm{sel}\left(\text{`c'},\mathit{FLExpr}\right)$}, \fbox{$\mathrm{sel}\left(\text{`t'},\mathit{FLExpr}\right)$}, \fbox{$\mathrm{sel}\left(\text{`e'},\mathit{FLExpr}\right)$}\right]\right)\right)$}, \fbox{$\mathrm{sel}\left(\text{`argument'},\fbox{$\mathrm{sel}\left(\text{`a'},str\right)$}\right)$}, \fbox{$\mathrm{sel}\left(\text{`literal'},\fbox{$\mathrm{sel}\left(\text{`i'},int\right)$}\right)$}\right]\right)\right)$
\item \textbf{anonymize-deanonymize}\\$\mathrm{p}\left(\text{`'},\mathit{FLFun},\mathrm{seq}\left(\left[\fbox{$\mathrm{sel}\left(\text{`f'},str\right)$}, \fbox{$\mathrm{sel}\left(\text{`args'},\star \left(str\right)\right)$}, \fbox{$\mathrm{sel}\left(\text{`body'},\mathit{FLExpr}\right)$}\right]\right)\right)$
\item \textbf{vertical-horizontal}  in $\mathit{FLExpr}$
\item \textbf{undefine-define}\\$\mathrm{p}\left(\text{`'},\mathit{FLOp},\varepsilon\right)$
\item \textbf{unlabel-designate}\\$\mathrm{p}\left(\fbox{\text{`fs'}},\mathit{FLPrg},\star \left(\mathit{FLFun}\right)\right)$
\item \textbf{extract-inline}  in $\mathit{FLExpr}$\\$\mathrm{p}\left(\text{`'},\mathit{FLExpr_1},\mathrm{seq}\left(\left[\mathit{FLExpr}, \mathit{FLOp}, \mathit{FLExpr}\right]\right)\right)$
\item \textbf{extract-inline}  in $\mathit{FLExpr}$\\$\mathrm{p}\left(\text{`'},\mathit{FLExpr_2},\mathrm{seq}\left(\left[str, \star \left(\mathit{FLExpr}\right)\right]\right)\right)$
\item \textbf{extract-inline}  in $\mathit{FLExpr}$\\$\mathrm{p}\left(\text{`'},\mathit{FLExpr_3},\mathrm{seq}\left(\left[\mathit{FLExpr}, \mathit{FLExpr}, \mathit{FLExpr}\right]\right)\right)$
\end{itemize}}

\section{Grammar in ANF}

\footnotesize\begin{center}\begin{tabular}{|l|c|}\hline
\multicolumn{1}{|>{\columncolor[gray]{.9}}c|}{\footnotesize \textbf{Production rule}} &
\multicolumn{1}{>{\columncolor[gray]{.9}}c|}{\footnotesize \textbf{Production signature}}
\\\hline
$\mathrm{p}\left(\text{`'},\mathit{FLPrg},\star \left(\mathit{FLFun}\right)\right)$	&	$\{ \langle \mathit{FLFun}, {*}\rangle\}$\\
$\mathrm{p}\left(\text{`'},\mathit{FLFun},\mathrm{seq}\left(\left[str, \star \left(str\right), \mathit{FLExpr}\right]\right)\right)$	&	$\{ \langle str, 1{*}\rangle, \langle \mathit{FLExpr}, 1\rangle\}$\\
$\mathrm{p}\left(\text{`'},\mathit{FLExpr},\mathit{FLExpr_1}\right)$	&	$\{ \langle \mathit{FLExpr_1}, 1\rangle\}$\\
$\mathrm{p}\left(\text{`'},\mathit{FLExpr},\mathit{FLExpr_2}\right)$	&	$\{ \langle \mathit{FLExpr_2}, 1\rangle\}$\\
$\mathrm{p}\left(\text{`'},\mathit{FLExpr},\mathit{FLExpr_3}\right)$	&	$\{ \langle \mathit{FLExpr_3}, 1\rangle\}$\\
$\mathrm{p}\left(\text{`'},\mathit{FLExpr},str\right)$	&	$\{ \langle str, 1\rangle\}$\\
$\mathrm{p}\left(\text{`'},\mathit{FLExpr},int\right)$	&	$\{ \langle int, 1\rangle\}$\\
$\mathrm{p}\left(\text{`'},\mathit{FLExpr_1},\mathrm{seq}\left(\left[\mathit{FLExpr}, \mathit{FLOp}, \mathit{FLExpr}\right]\right)\right)$	&	$\{ \langle \mathit{FLOp}, 1\rangle, \langle \mathit{FLExpr}, 11\rangle\}$\\
$\mathrm{p}\left(\text{`'},\mathit{FLExpr_2},\mathrm{seq}\left(\left[str, \star \left(\mathit{FLExpr}\right)\right]\right)\right)$	&	$\{ \langle str, 1\rangle, \langle \mathit{FLExpr}, {*}\rangle\}$\\
$\mathrm{p}\left(\text{`'},\mathit{FLExpr_3},\mathrm{seq}\left(\left[\mathit{FLExpr}, \mathit{FLExpr}, \mathit{FLExpr}\right]\right)\right)$	&	$\{ \langle \mathit{FLExpr}, 111\rangle\}$\\
\hline\end{tabular}\end{center}

\section{Nominal resolution}

Production rules are matched as follows (ANF on the left, master grammar on the right):
\begin{eqnarray*}
\mathrm{p}\left(\text{`'},\mathit{FLPrg},\star \left(\mathit{FLFun}\right)\right) & \Bumpeq & \mathrm{p}\left(\text{`'},\mathit{program},\plus \left(\mathit{function}\right)\right) \\
\mathrm{p}\left(\text{`'},\mathit{FLFun},\mathrm{seq}\left(\left[str, \star \left(str\right), \mathit{FLExpr}\right]\right)\right) & \Bumpeq & \mathrm{p}\left(\text{`'},\mathit{function},\mathrm{seq}\left(\left[str, \plus \left(str\right), \mathit{expression}\right]\right)\right) \\
\mathrm{p}\left(\text{`'},\mathit{FLExpr},\mathit{FLExpr_1}\right) & \bumpeq & \mathrm{p}\left(\text{`'},\mathit{expression},\mathit{binary}\right) \\
\mathrm{p}\left(\text{`'},\mathit{FLExpr},\mathit{FLExpr_2}\right) & \bumpeq & \mathrm{p}\left(\text{`'},\mathit{expression},\mathit{apply}\right) \\
\mathrm{p}\left(\text{`'},\mathit{FLExpr},\mathit{FLExpr_3}\right) & \bumpeq & \mathrm{p}\left(\text{`'},\mathit{expression},\mathit{conditional}\right) \\
\mathrm{p}\left(\text{`'},\mathit{FLExpr},str\right) & \bumpeq & \mathrm{p}\left(\text{`'},\mathit{expression},str\right) \\
\mathrm{p}\left(\text{`'},\mathit{FLExpr},int\right) & \bumpeq & \mathrm{p}\left(\text{`'},\mathit{expression},int\right) \\
\mathrm{p}\left(\text{`'},\mathit{FLExpr_1},\mathrm{seq}\left(\left[\mathit{FLExpr}, \mathit{FLOp}, \mathit{FLExpr}\right]\right)\right) & \bumpeq & \mathrm{p}\left(\text{`'},\mathit{binary},\mathrm{seq}\left(\left[\mathit{expression}, \mathit{operator}, \mathit{expression}\right]\right)\right) \\
\mathrm{p}\left(\text{`'},\mathit{FLExpr_2},\mathrm{seq}\left(\left[str, \star \left(\mathit{FLExpr}\right)\right]\right)\right) & \Bumpeq & \mathrm{p}\left(\text{`'},\mathit{apply},\mathrm{seq}\left(\left[str, \plus \left(\mathit{expression}\right)\right]\right)\right) \\
\mathrm{p}\left(\text{`'},\mathit{FLExpr_3},\mathrm{seq}\left(\left[\mathit{FLExpr}, \mathit{FLExpr}, \mathit{FLExpr}\right]\right)\right) & \bumpeq & \mathrm{p}\left(\text{`'},\mathit{conditional},\mathrm{seq}\left(\left[\mathit{expression}, \mathit{expression}, \mathit{expression}\right]\right)\right) \\
\end{eqnarray*}
This yields the following nominal mapping:
\begin{align*}\mathit{rascal-a} \:\diamond\: \mathit{master} =\:& \{\langle \mathit{FLFun},\mathit{function}\rangle,\\
 & \langle \mathit{FLExpr_2},\mathit{apply}\rangle,\\
 & \langle \mathit{FLPrg},\mathit{program}\rangle,\\
 & \langle \mathit{FLExpr},\mathit{expression}\rangle,\\
 & \langle int,int\rangle,\\
 & \langle str,str\rangle,\\
 & \langle \mathit{FLExpr_3},\mathit{conditional}\rangle,\\
 & \langle \mathit{FLOp},\mathit{operator}\rangle,\\
 & \langle \mathit{FLExpr_1},\mathit{binary}\rangle\}\end{align*}
 Which is exercised with these grammar transformation steps:

{\footnotesize\begin{itemize}
\item \textbf{renameN-renameN} $\mathit{FLFun}$ to $\mathit{function}$
\item \textbf{renameN-renameN} $\mathit{FLExpr_2}$ to $\mathit{apply}$
\item \textbf{renameN-renameN} $\mathit{FLPrg}$ to $\mathit{program}$
\item \textbf{renameN-renameN} $\mathit{FLExpr}$ to $\mathit{expression}$
\item \textbf{renameN-renameN} $\mathit{FLExpr_3}$ to $\mathit{conditional}$
\item \textbf{renameN-renameN} $\mathit{FLOp}$ to $\mathit{operator}$
\item \textbf{renameN-renameN} $\mathit{FLExpr_1}$ to $\mathit{binary}$
\end{itemize}}

\section{Structural resolution}
{\footnotesize\begin{itemize}
\item \textbf{narrow-widen}  in $\mathit{program}$\\$\star \left(\mathit{function}\right)$\\$\plus \left(\mathit{function}\right)$
\item \textbf{narrow-widen}  in $\mathit{function}$\\$\star \left(str\right)$\\$\plus \left(str\right)$
\item \textbf{narrow-widen}  in $\mathit{apply}$\\$\star \left(\mathit{expression}\right)$\\$\plus \left(\mathit{expression}\right)$
\end{itemize}}

\chapter{Rascal Concrete Syntax Definition}

 Source name: \textbf{rascal-c}

\section{Source grammar}

\begin{itemize}\item Source artifact: \href{http://github.com/grammarware/slps/blob/master/topics/fl/rascal/Concrete.rsc}{topics/fl/rascal/Concrete.rsc}\item Grammar extractor: \href{http://github.com/grammarware/slps/blob/master/shared/rascal/src/extract/RascalSyntax2BGF.rsc}{shared/rascal/src/extract/RascalSyntax2BGF.rsc}\end{itemize}

\footnotesize\begin{center}\begin{tabular}{|l|}\hline
\multicolumn{1}{|>{\columncolor[gray]{.9}}c|}{\footnotesize \textbf{Production rules}}
\\\hline
$\mathrm{p}(\text{`prg'},\mathit{Program},\mathrm{sel}\left(\text{`functions'},\mathrm{s}{+}\left(\mathit{Function},{\swarrow}\right)\right))$	\\
$\mathrm{p}(\text{`ifThenElse'},\mathit{Expr},\mathrm{seq}\left(\left[\text{`if'}, \mathrm{sel}\left(\text{`cond'},\mathit{Expr}\right), \text{`then'}, \mathrm{sel}\left(\text{`thenbranch'},\mathit{Expr}\right), \text{`else'}, \mathrm{sel}\left(\text{`elsebranch'},\mathit{Expr}\right)\right]\right))$	\\
$\mathrm{p}(\text{`'},\mathit{Expr},\mathrm{seq}\left(\left[\text{`('}, \mathrm{sel}\left(\text{`e'},\mathit{Expr}\right), \text{`)'}\right]\right))$	\\
$\mathrm{p}(\text{`literal'},\mathit{Expr},\mathrm{sel}\left(\text{`i'},\mathit{Int}\right))$	\\
$\mathrm{p}(\text{`argument'},\mathit{Expr},\mathrm{sel}\left(\text{`a'},\mathit{Name}\right))$	\\
$\mathrm{p}(\text{`binary'},\mathit{Expr},\mathrm{seq}\left(\left[\mathrm{sel}\left(\text{`lexpr'},\mathit{Expr}\right), \mathrm{sel}\left(\text{`op'},\mathit{Ops}\right), \mathrm{sel}\left(\text{`rexpr'},\mathit{Expr}\right)\right]\right))$	\\
$\mathrm{p}(\text{`apply'},\mathit{Expr},\mathrm{seq}\left(\left[\mathrm{sel}\left(\text{`f'},\mathit{Name}\right), \mathrm{sel}\left(\text{`vargs'},\plus \left(\mathit{Expr}\right)\right)\right]\right))$	\\
$\mathrm{p}(\text{`plus'},\mathit{Ops},\text{`+'})$	\\
$\mathrm{p}(\text{`equal'},\mathit{Ops},\text{`=='})$	\\
$\mathrm{p}(\text{`minus'},\mathit{Ops},\text{`-'})$	\\
$\mathrm{p}(\text{`fun'},\mathit{Function},\mathrm{seq}\left(\left[\mathrm{sel}\left(\text{`f'},\mathit{Name}\right), \mathrm{sel}\left(\text{`args'},\plus \left(\mathit{Name}\right)\right), \text{`='}, \mathrm{sel}\left(\text{`body'},\mathit{Expr}\right)\right]\right))$	\\
\hline\end{tabular}\end{center}

\section{Normalizations}
{\footnotesize\begin{itemize}
\item \textbf{reroot-reroot} $\left[\right]$ to $\left[\mathit{Program}\right]$
\item \textbf{unlabel-designate}\\$\mathrm{p}\left(\fbox{\text{`prg'}},\mathit{Program},\mathrm{sel}\left(\text{`functions'},\mathrm{s}{+}\left(\mathit{Function},{\swarrow}\right)\right)\right)$
\item \textbf{unlabel-designate}\\$\mathrm{p}\left(\fbox{\text{`ifThenElse'}},\mathit{Expr},\mathrm{seq}\left(\left[\text{`if'}, \mathrm{sel}\left(\text{`cond'},\mathit{Expr}\right), \text{`then'}, \mathrm{sel}\left(\text{`thenbranch'},\mathit{Expr}\right), \text{`else'}, \mathrm{sel}\left(\text{`elsebranch'},\mathit{Expr}\right)\right]\right)\right)$
\item \textbf{unlabel-designate}\\$\mathrm{p}\left(\fbox{\text{`literal'}},\mathit{Expr},\mathrm{sel}\left(\text{`i'},\mathit{Int}\right)\right)$
\item \textbf{unlabel-designate}\\$\mathrm{p}\left(\fbox{\text{`argument'}},\mathit{Expr},\mathrm{sel}\left(\text{`a'},\mathit{Name}\right)\right)$
\item \textbf{unlabel-designate}\\$\mathrm{p}\left(\fbox{\text{`binary'}},\mathit{Expr},\mathrm{seq}\left(\left[\mathrm{sel}\left(\text{`lexpr'},\mathit{Expr}\right), \mathrm{sel}\left(\text{`op'},\mathit{Ops}\right), \mathrm{sel}\left(\text{`rexpr'},\mathit{Expr}\right)\right]\right)\right)$
\item \textbf{unlabel-designate}\\$\mathrm{p}\left(\fbox{\text{`apply'}},\mathit{Expr},\mathrm{seq}\left(\left[\mathrm{sel}\left(\text{`f'},\mathit{Name}\right), \mathrm{sel}\left(\text{`vargs'},\plus \left(\mathit{Expr}\right)\right)\right]\right)\right)$
\item \textbf{unlabel-designate}\\$\mathrm{p}\left(\fbox{\text{`plus'}},\mathit{Ops},\text{`+'}\right)$
\item \textbf{unlabel-designate}\\$\mathrm{p}\left(\fbox{\text{`equal'}},\mathit{Ops},\text{`=='}\right)$
\item \textbf{unlabel-designate}\\$\mathrm{p}\left(\fbox{\text{`minus'}},\mathit{Ops},\text{`-'}\right)$
\item \textbf{unlabel-designate}\\$\mathrm{p}\left(\fbox{\text{`fun'}},\mathit{Function},\mathrm{seq}\left(\left[\mathrm{sel}\left(\text{`f'},\mathit{Name}\right), \mathrm{sel}\left(\text{`args'},\plus \left(\mathit{Name}\right)\right), \text{`='}, \mathrm{sel}\left(\text{`body'},\mathit{Expr}\right)\right]\right)\right)$
\item \textbf{anonymize-deanonymize}\\$\mathrm{p}\left(\text{`'},\mathit{Expr},\mathrm{seq}\left(\left[\fbox{$\mathrm{sel}\left(\text{`f'},\mathit{Name}\right)$}, \fbox{$\mathrm{sel}\left(\text{`vargs'},\plus \left(\mathit{Expr}\right)\right)$}\right]\right)\right)$
\item \textbf{anonymize-deanonymize}\\$\mathrm{p}\left(\text{`'},\mathit{Function},\mathrm{seq}\left(\left[\fbox{$\mathrm{sel}\left(\text{`f'},\mathit{Name}\right)$}, \fbox{$\mathrm{sel}\left(\text{`args'},\plus \left(\mathit{Name}\right)\right)$}, \text{`='}, \fbox{$\mathrm{sel}\left(\text{`body'},\mathit{Expr}\right)$}\right]\right)\right)$
\item \textbf{anonymize-deanonymize}\\$\mathrm{p}\left(\text{`'},\mathit{Expr},\mathrm{seq}\left(\left[\text{`if'}, \fbox{$\mathrm{sel}\left(\text{`cond'},\mathit{Expr}\right)$}, \text{`then'}, \fbox{$\mathrm{sel}\left(\text{`thenbranch'},\mathit{Expr}\right)$}, \text{`else'}, \fbox{$\mathrm{sel}\left(\text{`elsebranch'},\mathit{Expr}\right)$}\right]\right)\right)$
\item \textbf{anonymize-deanonymize}\\$\mathrm{p}\left(\text{`'},\mathit{Expr},\mathrm{seq}\left(\left[\fbox{$\mathrm{sel}\left(\text{`lexpr'},\mathit{Expr}\right)$}, \fbox{$\mathrm{sel}\left(\text{`op'},\mathit{Ops}\right)$}, \fbox{$\mathrm{sel}\left(\text{`rexpr'},\mathit{Expr}\right)$}\right]\right)\right)$
\item \textbf{anonymize-deanonymize}\\$\mathrm{p}\left(\text{`'},\mathit{Expr},\mathrm{seq}\left(\left[\text{`('}, \fbox{$\mathrm{sel}\left(\text{`e'},\mathit{Expr}\right)$}, \text{`)'}\right]\right)\right)$
\item \textbf{abstractize-concretize}\\$\mathrm{p}\left(\text{`'},\mathit{Expr},\mathrm{seq}\left(\left[\fbox{$\text{`('}$}, \mathit{Expr}, \fbox{$\text{`)'}$}\right]\right)\right)$
\item \textbf{abstractize-concretize}\\$\mathrm{p}\left(\text{`'},\mathit{Function},\mathrm{seq}\left(\left[\mathit{Name}, \plus \left(\mathit{Name}\right), \fbox{$\text{`='}$}, \mathit{Expr}\right]\right)\right)$
\item \textbf{abstractize-concretize}\\$\mathrm{p}\left(\text{`'},\mathit{Ops},\fbox{$\text{`-'}$}\right)$
\item \textbf{abstractize-concretize}\\$\mathrm{p}\left(\text{`'},\mathit{Ops},\fbox{$\text{`+'}$}\right)$
\item \textbf{abstractize-concretize}\\$\mathrm{p}\left(\text{`'},\mathit{Ops},\fbox{$\text{`=='}$}\right)$
\item \textbf{abstractize-concretize}\\$\mathrm{p}\left(\text{`functions'},\mathit{Program},\mathrm{s}{+}\left(\mathit{Function},\fbox{${\swarrow}$}\right)\right)$
\item \textbf{abstractize-concretize}\\$\mathrm{p}\left(\text{`'},\mathit{Expr},\mathrm{seq}\left(\left[\fbox{$\text{`if'}$}, \mathit{Expr}, \fbox{$\text{`then'}$}, \mathit{Expr}, \fbox{$\text{`else'}$}, \mathit{Expr}\right]\right)\right)$
\item \textbf{undefine-define}\\$\mathrm{p}\left(\text{`'},\mathit{Ops},\varepsilon\right)$
\item \textbf{abridge-detour}\\$\mathrm{p}\left(\text{`'},\mathit{Expr},\mathit{Expr}\right)$
\item \textbf{unlabel-designate}\\$\mathrm{p}\left(\fbox{\text{`functions'}},\mathit{Program},\plus \left(\mathit{Function}\right)\right)$
\item \textbf{unlabel-designate}\\$\mathrm{p}\left(\fbox{\text{`i'}},\mathit{Expr},\mathit{Int}\right)$
\item \textbf{unlabel-designate}\\$\mathrm{p}\left(\fbox{\text{`a'}},\mathit{Expr},\mathit{Name}\right)$
\item \textbf{extract-inline}  in $\mathit{Expr}$\\$\mathrm{p}\left(\text{`'},\mathit{Expr_1},\mathrm{seq}\left(\left[\mathit{Expr}, \mathit{Expr}, \mathit{Expr}\right]\right)\right)$
\item \textbf{extract-inline}  in $\mathit{Expr}$\\$\mathrm{p}\left(\text{`'},\mathit{Expr_2},\mathrm{seq}\left(\left[\mathit{Expr}, \mathit{Ops}, \mathit{Expr}\right]\right)\right)$
\item \textbf{extract-inline}  in $\mathit{Expr}$\\$\mathrm{p}\left(\text{`'},\mathit{Expr_3},\mathrm{seq}\left(\left[\mathit{Name}, \plus \left(\mathit{Expr}\right)\right]\right)\right)$
\end{itemize}}

\section{Grammar in ANF}

\footnotesize\begin{center}\begin{tabular}{|l|c|}\hline
\multicolumn{1}{|>{\columncolor[gray]{.9}}c|}{\footnotesize \textbf{Production rule}} &
\multicolumn{1}{>{\columncolor[gray]{.9}}c|}{\footnotesize \textbf{Production signature}}
\\\hline
$\mathrm{p}\left(\text{`'},\mathit{Program},\plus \left(\mathit{Function}\right)\right)$	&	$\{ \langle \mathit{Function}, {+}\rangle\}$\\
$\mathrm{p}\left(\text{`'},\mathit{Expr},\mathit{Expr_1}\right)$	&	$\{ \langle \mathit{Expr_1}, 1\rangle\}$\\
$\mathrm{p}\left(\text{`'},\mathit{Expr},\mathit{Int}\right)$	&	$\{ \langle \mathit{Int}, 1\rangle\}$\\
$\mathrm{p}\left(\text{`'},\mathit{Expr},\mathit{Name}\right)$	&	$\{ \langle \mathit{Name}, 1\rangle\}$\\
$\mathrm{p}\left(\text{`'},\mathit{Expr},\mathit{Expr_2}\right)$	&	$\{ \langle \mathit{Expr_2}, 1\rangle\}$\\
$\mathrm{p}\left(\text{`'},\mathit{Expr},\mathit{Expr_3}\right)$	&	$\{ \langle \mathit{Expr_3}, 1\rangle\}$\\
$\mathrm{p}\left(\text{`'},\mathit{Function},\mathrm{seq}\left(\left[\mathit{Name}, \plus \left(\mathit{Name}\right), \mathit{Expr}\right]\right)\right)$	&	$\{ \langle \mathit{Expr}, 1\rangle, \langle \mathit{Name}, 1{+}\rangle\}$\\
$\mathrm{p}\left(\text{`'},\mathit{Expr_1},\mathrm{seq}\left(\left[\mathit{Expr}, \mathit{Expr}, \mathit{Expr}\right]\right)\right)$	&	$\{ \langle \mathit{Expr}, 111\rangle\}$\\
$\mathrm{p}\left(\text{`'},\mathit{Expr_2},\mathrm{seq}\left(\left[\mathit{Expr}, \mathit{Ops}, \mathit{Expr}\right]\right)\right)$	&	$\{ \langle \mathit{Ops}, 1\rangle, \langle \mathit{Expr}, 11\rangle\}$\\
$\mathrm{p}\left(\text{`'},\mathit{Expr_3},\mathrm{seq}\left(\left[\mathit{Name}, \plus \left(\mathit{Expr}\right)\right]\right)\right)$	&	$\{ \langle \mathit{Expr}, {+}\rangle, \langle \mathit{Name}, 1\rangle\}$\\
\hline\end{tabular}\end{center}

\section{Nominal resolution}

Production rules are matched as follows (ANF on the left, master grammar on the right):
\begin{eqnarray*}
\mathrm{p}\left(\text{`'},\mathit{Program},\plus \left(\mathit{Function}\right)\right) & \bumpeq & \mathrm{p}\left(\text{`'},\mathit{program},\plus \left(\mathit{function}\right)\right) \\
\mathrm{p}\left(\text{`'},\mathit{Expr},\mathit{Expr_1}\right) & \bumpeq & \mathrm{p}\left(\text{`'},\mathit{expression},\mathit{conditional}\right) \\
\mathrm{p}\left(\text{`'},\mathit{Expr},\mathit{Int}\right) & \bumpeq & \mathrm{p}\left(\text{`'},\mathit{expression},int\right) \\
\mathrm{p}\left(\text{`'},\mathit{Expr},\mathit{Name}\right) & \bumpeq & \mathrm{p}\left(\text{`'},\mathit{expression},str\right) \\
\mathrm{p}\left(\text{`'},\mathit{Expr},\mathit{Expr_2}\right) & \bumpeq & \mathrm{p}\left(\text{`'},\mathit{expression},\mathit{binary}\right) \\
\mathrm{p}\left(\text{`'},\mathit{Expr},\mathit{Expr_3}\right) & \bumpeq & \mathrm{p}\left(\text{`'},\mathit{expression},\mathit{apply}\right) \\
\mathrm{p}\left(\text{`'},\mathit{Function},\mathrm{seq}\left(\left[\mathit{Name}, \plus \left(\mathit{Name}\right), \mathit{Expr}\right]\right)\right) & \bumpeq & \mathrm{p}\left(\text{`'},\mathit{function},\mathrm{seq}\left(\left[str, \plus \left(str\right), \mathit{expression}\right]\right)\right) \\
\mathrm{p}\left(\text{`'},\mathit{Expr_1},\mathrm{seq}\left(\left[\mathit{Expr}, \mathit{Expr}, \mathit{Expr}\right]\right)\right) & \bumpeq & \mathrm{p}\left(\text{`'},\mathit{conditional},\mathrm{seq}\left(\left[\mathit{expression}, \mathit{expression}, \mathit{expression}\right]\right)\right) \\
\mathrm{p}\left(\text{`'},\mathit{Expr_2},\mathrm{seq}\left(\left[\mathit{Expr}, \mathit{Ops}, \mathit{Expr}\right]\right)\right) & \bumpeq & \mathrm{p}\left(\text{`'},\mathit{binary},\mathrm{seq}\left(\left[\mathit{expression}, \mathit{operator}, \mathit{expression}\right]\right)\right) \\
\mathrm{p}\left(\text{`'},\mathit{Expr_3},\mathrm{seq}\left(\left[\mathit{Name}, \plus \left(\mathit{Expr}\right)\right]\right)\right) & \bumpeq & \mathrm{p}\left(\text{`'},\mathit{apply},\mathrm{seq}\left(\left[str, \plus \left(\mathit{expression}\right)\right]\right)\right) \\
\end{eqnarray*}
This yields the following nominal mapping:
\begin{align*}\mathit{rascal-c} \:\diamond\: \mathit{master} =\:& \{\langle \mathit{Expr_2},\mathit{binary}\rangle,\\
 & \langle \mathit{Int},int\rangle,\\
 & \langle \mathit{Expr_1},\mathit{conditional}\rangle,\\
 & \langle \mathit{Function},\mathit{function}\rangle,\\
 & \langle \mathit{Program},\mathit{program}\rangle,\\
 & \langle \mathit{Name},str\rangle,\\
 & \langle \mathit{Expr_3},\mathit{apply}\rangle,\\
 & \langle \mathit{Expr},\mathit{expression}\rangle,\\
 & \langle \mathit{Ops},\mathit{operator}\rangle\}\end{align*}
 Which is exercised with these grammar transformation steps:

{\footnotesize\begin{itemize}
\item \textbf{renameN-renameN} $\mathit{Expr_2}$ to $\mathit{binary}$
\item \textbf{renameN-renameN} $\mathit{Int}$ to $int$
\item \textbf{renameN-renameN} $\mathit{Expr_1}$ to $\mathit{conditional}$
\item \textbf{renameN-renameN} $\mathit{Function}$ to $\mathit{function}$
\item \textbf{renameN-renameN} $\mathit{Program}$ to $\mathit{program}$
\item \textbf{renameN-renameN} $\mathit{Name}$ to $str$
\item \textbf{renameN-renameN} $\mathit{Expr_3}$ to $\mathit{apply}$
\item \textbf{renameN-renameN} $\mathit{Expr}$ to $\mathit{expression}$
\item \textbf{renameN-renameN} $\mathit{Ops}$ to $\mathit{operator}$
\end{itemize}}

\chapter{Syntax Definition Formalism}

 Source name: \textbf{sdf}

\section{Source grammar}

\begin{itemize}\item Source artifact: \href{http://github.com/grammarware/slps/blob/master/topics/fl/asfsdf/Syntax.sdf}{topics/fl/asfsdf/Syntax.sdf}\item Grammar extractor: \href{http://github.com/grammarware/slps/blob/master/topics/extraction/sdf/Main.sdf}{topics/extraction/sdf/Main.sdf}\item Grammar extractor: \href{http://github.com/grammarware/slps/blob/master/topics/extraction/sdf/Main.asf}{topics/extraction/sdf/Main.asf}\item Grammar extractor: \href{http://github.com/grammarware/slps/blob/master/topics/extraction/sdf/Tokens.sdf}{topics/extraction/sdf/Tokens.sdf}\item Grammar extractor: \href{http://github.com/grammarware/slps/blob/master/topics/extraction/sdf/Tokens.asf}{topics/extraction/sdf/Tokens.asf}\end{itemize}

\footnotesize\begin{center}\begin{tabular}{|l|}\hline
\multicolumn{1}{|>{\columncolor[gray]{.9}}c|}{\footnotesize \textbf{Production rules}}
\\\hline
$\mathrm{p}(\text{`'},\mathit{Program},\plus \left(\mathit{Function}\right))$	\\
$\mathrm{p}(\text{`'},\mathit{Function},\mathrm{seq}\left(\left[\mathit{Name}, \plus \left(\mathit{Name}\right), \text{`='}, \mathit{Expr}, \plus \left(\mathit{Newline}\right)\right]\right))$	\\
$\mathrm{p}(\text{`binary'},\mathit{Expr},\mathrm{seq}\left(\left[\mathit{Expr}, \mathit{Ops}, \mathit{Expr}\right]\right))$	\\
$\mathrm{p}(\text{`apply'},\mathit{Expr},\mathrm{seq}\left(\left[\mathit{Name}, \plus \left(\mathit{Expr}\right)\right]\right))$	\\
$\mathrm{p}(\text{`ifThenElse'},\mathit{Expr},\mathrm{seq}\left(\left[\text{`if'}, \mathit{Expr}, \text{`then'}, \mathit{Expr}, \text{`else'}, \mathit{Expr}\right]\right))$	\\
$\mathrm{p}(\text{`'},\mathit{Expr},\mathrm{seq}\left(\left[\text{`('}, \mathit{Expr}, \text{`)'}\right]\right))$	\\
$\mathrm{p}(\text{`argument'},\mathit{Expr},\mathit{Name})$	\\
$\mathrm{p}(\text{`literal'},\mathit{Expr},\mathit{Int})$	\\
$\mathrm{p}(\text{`minus'},\mathit{Ops},\text{`-'})$	\\
$\mathrm{p}(\text{`plus'},\mathit{Ops},\text{`+'})$	\\
$\mathrm{p}(\text{`equal'},\mathit{Ops},\text{`=='})$	\\
\hline\end{tabular}\end{center}

\section{Normalizations}
{\footnotesize\begin{itemize}
\item \textbf{reroot-reroot} $\left[\right]$ to $\left[\mathit{Program}\right]$
\item \textbf{unlabel-designate}\\$\mathrm{p}\left(\fbox{\text{`binary'}},\mathit{Expr},\mathrm{seq}\left(\left[\mathit{Expr}, \mathit{Ops}, \mathit{Expr}\right]\right)\right)$
\item \textbf{unlabel-designate}\\$\mathrm{p}\left(\fbox{\text{`apply'}},\mathit{Expr},\mathrm{seq}\left(\left[\mathit{Name}, \plus \left(\mathit{Expr}\right)\right]\right)\right)$
\item \textbf{unlabel-designate}\\$\mathrm{p}\left(\fbox{\text{`ifThenElse'}},\mathit{Expr},\mathrm{seq}\left(\left[\text{`if'}, \mathit{Expr}, \text{`then'}, \mathit{Expr}, \text{`else'}, \mathit{Expr}\right]\right)\right)$
\item \textbf{unlabel-designate}\\$\mathrm{p}\left(\fbox{\text{`argument'}},\mathit{Expr},\mathit{Name}\right)$
\item \textbf{unlabel-designate}\\$\mathrm{p}\left(\fbox{\text{`literal'}},\mathit{Expr},\mathit{Int}\right)$
\item \textbf{unlabel-designate}\\$\mathrm{p}\left(\fbox{\text{`minus'}},\mathit{Ops},\text{`-'}\right)$
\item \textbf{unlabel-designate}\\$\mathrm{p}\left(\fbox{\text{`plus'}},\mathit{Ops},\text{`+'}\right)$
\item \textbf{unlabel-designate}\\$\mathrm{p}\left(\fbox{\text{`equal'}},\mathit{Ops},\text{`=='}\right)$
\item \textbf{abstractize-concretize}\\$\mathrm{p}\left(\text{`'},\mathit{Expr},\mathrm{seq}\left(\left[\fbox{$\text{`('}$}, \mathit{Expr}, \fbox{$\text{`)'}$}\right]\right)\right)$
\item \textbf{abstractize-concretize}\\$\mathrm{p}\left(\text{`'},\mathit{Ops},\fbox{$\text{`+'}$}\right)$
\item \textbf{abstractize-concretize}\\$\mathrm{p}\left(\text{`'},\mathit{Ops},\fbox{$\text{`-'}$}\right)$
\item \textbf{abstractize-concretize}\\$\mathrm{p}\left(\text{`'},\mathit{Ops},\fbox{$\text{`=='}$}\right)$
\item \textbf{abstractize-concretize}\\$\mathrm{p}\left(\text{`'},\mathit{Expr},\mathrm{seq}\left(\left[\fbox{$\text{`if'}$}, \mathit{Expr}, \fbox{$\text{`then'}$}, \mathit{Expr}, \fbox{$\text{`else'}$}, \mathit{Expr}\right]\right)\right)$
\item \textbf{abstractize-concretize}\\$\mathrm{p}\left(\text{`'},\mathit{Function},\mathrm{seq}\left(\left[\mathit{Name}, \plus \left(\mathit{Name}\right), \fbox{$\text{`='}$}, \mathit{Expr}, \plus \left(\mathit{Newline}\right)\right]\right)\right)$
\item \textbf{undefine-define}\\$\mathrm{p}\left(\text{`'},\mathit{Ops},\varepsilon\right)$
\item \textbf{abridge-detour}\\$\mathrm{p}\left(\text{`'},\mathit{Expr},\mathit{Expr}\right)$
\item \textbf{extract-inline}  in $\mathit{Expr}$\\$\mathrm{p}\left(\text{`'},\mathit{Expr_1},\mathrm{seq}\left(\left[\mathit{Expr}, \mathit{Ops}, \mathit{Expr}\right]\right)\right)$
\item \textbf{extract-inline}  in $\mathit{Expr}$\\$\mathrm{p}\left(\text{`'},\mathit{Expr_2},\mathrm{seq}\left(\left[\mathit{Name}, \plus \left(\mathit{Expr}\right)\right]\right)\right)$
\item \textbf{extract-inline}  in $\mathit{Expr}$\\$\mathrm{p}\left(\text{`'},\mathit{Expr_3},\mathrm{seq}\left(\left[\mathit{Expr}, \mathit{Expr}, \mathit{Expr}\right]\right)\right)$
\end{itemize}}

\section{Grammar in ANF}

\footnotesize\begin{center}\begin{tabular}{|l|c|}\hline
\multicolumn{1}{|>{\columncolor[gray]{.9}}c|}{\footnotesize \textbf{Production rule}} &
\multicolumn{1}{>{\columncolor[gray]{.9}}c|}{\footnotesize \textbf{Production signature}}
\\\hline
$\mathrm{p}\left(\text{`'},\mathit{Program},\plus \left(\mathit{Function}\right)\right)$	&	$\{ \langle \mathit{Function}, {+}\rangle\}$\\
$\mathrm{p}\left(\text{`'},\mathit{Function},\mathrm{seq}\left(\left[\mathit{Name}, \plus \left(\mathit{Name}\right), \mathit{Expr}, \plus \left(\mathit{Newline}\right)\right]\right)\right)$	&	$\{ \langle \mathit{Expr}, 1\rangle, \langle \mathit{Newline}, {+}\rangle, \langle \mathit{Name}, 1{+}\rangle\}$\\
$\mathrm{p}\left(\text{`'},\mathit{Expr},\mathit{Expr_1}\right)$	&	$\{ \langle \mathit{Expr_1}, 1\rangle\}$\\
$\mathrm{p}\left(\text{`'},\mathit{Expr},\mathit{Expr_2}\right)$	&	$\{ \langle \mathit{Expr_2}, 1\rangle\}$\\
$\mathrm{p}\left(\text{`'},\mathit{Expr},\mathit{Expr_3}\right)$	&	$\{ \langle \mathit{Expr_3}, 1\rangle\}$\\
$\mathrm{p}\left(\text{`'},\mathit{Expr},\mathit{Name}\right)$	&	$\{ \langle \mathit{Name}, 1\rangle\}$\\
$\mathrm{p}\left(\text{`'},\mathit{Expr},\mathit{Int}\right)$	&	$\{ \langle \mathit{Int}, 1\rangle\}$\\
$\mathrm{p}\left(\text{`'},\mathit{Expr_1},\mathrm{seq}\left(\left[\mathit{Expr}, \mathit{Ops}, \mathit{Expr}\right]\right)\right)$	&	$\{ \langle \mathit{Ops}, 1\rangle, \langle \mathit{Expr}, 11\rangle\}$\\
$\mathrm{p}\left(\text{`'},\mathit{Expr_2},\mathrm{seq}\left(\left[\mathit{Name}, \plus \left(\mathit{Expr}\right)\right]\right)\right)$	&	$\{ \langle \mathit{Expr}, {+}\rangle, \langle \mathit{Name}, 1\rangle\}$\\
$\mathrm{p}\left(\text{`'},\mathit{Expr_3},\mathrm{seq}\left(\left[\mathit{Expr}, \mathit{Expr}, \mathit{Expr}\right]\right)\right)$	&	$\{ \langle \mathit{Expr}, 111\rangle\}$\\
\hline\end{tabular}\end{center}

\section{Nominal resolution}

Production rules are matched as follows (ANF on the left, master grammar on the right):
\begin{eqnarray*}
\mathrm{p}\left(\text{`'},\mathit{Program},\plus \left(\mathit{Function}\right)\right) & \bumpeq & \mathrm{p}\left(\text{`'},\mathit{program},\plus \left(\mathit{function}\right)\right) \\
\mathrm{p}\left(\text{`'},\mathit{Function},\mathrm{seq}\left(\left[\mathit{Name}, \plus \left(\mathit{Name}\right), \mathit{Expr}, \plus \left(\mathit{Newline}\right)\right]\right)\right) & \Bumpeq & \mathrm{p}\left(\text{`'},\mathit{function},\mathrm{seq}\left(\left[str, \plus \left(str\right), \mathit{expression}\right]\right)\right) \\
\mathrm{p}\left(\text{`'},\mathit{Expr},\mathit{Expr_1}\right) & \bumpeq & \mathrm{p}\left(\text{`'},\mathit{expression},\mathit{binary}\right) \\
\mathrm{p}\left(\text{`'},\mathit{Expr},\mathit{Expr_2}\right) & \bumpeq & \mathrm{p}\left(\text{`'},\mathit{expression},\mathit{apply}\right) \\
\mathrm{p}\left(\text{`'},\mathit{Expr},\mathit{Expr_3}\right) & \bumpeq & \mathrm{p}\left(\text{`'},\mathit{expression},\mathit{conditional}\right) \\
\mathrm{p}\left(\text{`'},\mathit{Expr},\mathit{Name}\right) & \bumpeq & \mathrm{p}\left(\text{`'},\mathit{expression},str\right) \\
\mathrm{p}\left(\text{`'},\mathit{Expr},\mathit{Int}\right) & \bumpeq & \mathrm{p}\left(\text{`'},\mathit{expression},int\right) \\
\mathrm{p}\left(\text{`'},\mathit{Expr_1},\mathrm{seq}\left(\left[\mathit{Expr}, \mathit{Ops}, \mathit{Expr}\right]\right)\right) & \bumpeq & \mathrm{p}\left(\text{`'},\mathit{binary},\mathrm{seq}\left(\left[\mathit{expression}, \mathit{operator}, \mathit{expression}\right]\right)\right) \\
\mathrm{p}\left(\text{`'},\mathit{Expr_2},\mathrm{seq}\left(\left[\mathit{Name}, \plus \left(\mathit{Expr}\right)\right]\right)\right) & \bumpeq & \mathrm{p}\left(\text{`'},\mathit{apply},\mathrm{seq}\left(\left[str, \plus \left(\mathit{expression}\right)\right]\right)\right) \\
\mathrm{p}\left(\text{`'},\mathit{Expr_3},\mathrm{seq}\left(\left[\mathit{Expr}, \mathit{Expr}, \mathit{Expr}\right]\right)\right) & \bumpeq & \mathrm{p}\left(\text{`'},\mathit{conditional},\mathrm{seq}\left(\left[\mathit{expression}, \mathit{expression}, \mathit{expression}\right]\right)\right) \\
\end{eqnarray*}
This yields the following nominal mapping:
\begin{align*}\mathit{sdf} \:\diamond\: \mathit{master} =\:& \{\langle \mathit{Expr_3},\mathit{conditional}\rangle,\\
 & \langle \mathit{Int},int\rangle,\\
 & \langle \mathit{Expr_1},\mathit{binary}\rangle,\\
 & \langle \mathit{Newline},\omega\rangle,\\
 & \langle \mathit{Function},\mathit{function}\rangle,\\
 & \langle \mathit{Program},\mathit{program}\rangle,\\
 & \langle \mathit{Name},str\rangle,\\
 & \langle \mathit{Expr},\mathit{expression}\rangle,\\
 & \langle \mathit{Ops},\mathit{operator}\rangle,\\
 & \langle \mathit{Expr_2},\mathit{apply}\rangle\}\end{align*}
 Which is exercised with these grammar transformation steps:

{\footnotesize\begin{itemize}
\item \textbf{renameN-renameN} $\mathit{Expr_3}$ to $\mathit{conditional}$
\item \textbf{renameN-renameN} $\mathit{Int}$ to $int$
\item \textbf{renameN-renameN} $\mathit{Expr_1}$ to $\mathit{binary}$
\item \textbf{renameN-renameN} $\mathit{Function}$ to $\mathit{function}$
\item \textbf{renameN-renameN} $\mathit{Program}$ to $\mathit{program}$
\item \textbf{renameN-renameN} $\mathit{Name}$ to $str$
\item \textbf{renameN-renameN} $\mathit{Expr}$ to $\mathit{expression}$
\item \textbf{renameN-renameN} $\mathit{Ops}$ to $\mathit{operator}$
\item \textbf{renameN-renameN} $\mathit{Expr_2}$ to $\mathit{apply}$
\end{itemize}}

\section{Structural resolution}
{\footnotesize\begin{itemize}
\item \textbf{project-inject}\\$\mathrm{p}\left(\text{`'},\mathit{function},\mathrm{seq}\left(\left[str, \plus \left(str\right), \mathit{expression}, \plus \left(\fbox{$\mathit{Newline}$}\right)\right]\right)\right)$
\end{itemize}}

\chapter{TXL}

 Source name: \textbf{txl}

\section{Source grammar}

\begin{itemize}\item Source artifact: \href{http://github.com/grammarware/slps/blob/master/topics/fl/txl/FL.Txl}{topics/fl/txl/FL.Txl}\item Grammar extractor: \href{http://github.com/grammarware/slps/blob/master/topics/extraction/txl/txl2bgf.xslt}{topics/extraction/txl/txl2bgf.xslt}\end{itemize}

\footnotesize\begin{center}\begin{tabular}{|l|}\hline
\multicolumn{1}{|>{\columncolor[gray]{.9}}c|}{\footnotesize \textbf{Production rules}}
\\\hline
$\mathrm{p}(\text{`'},\mathit{program},\plus \left(\mathit{fun}\right))$	\\
$\mathrm{p}(\text{`'},\mathit{fun},\mathrm{seq}\left(\left[\mathit{id}, \plus \left(\mathit{id}\right), \text{`='}, \mathit{expression}, \mathit{newline}\right]\right))$	\\
$\mathrm{p}(\text{`'},\mathit{expression},\mathrm{choice}([\mathrm{seq}\left(\left[\mathit{expression}, \mathit{op}, \mathit{expression}\right]\right),$\\$\qquad\qquad\mathrm{seq}\left(\left[\mathit{id}, \plus \left(\mathit{expression}\right)\right]\right),$\\$\qquad\qquad\mathrm{seq}\left(\left[\text{`if'}, \mathit{expression}, \text{`then'}, \mathit{expression}, \text{`else'}, \mathit{expression}\right]\right),$\\$\qquad\qquad\mathrm{seq}\left(\left[\text{`('}, \mathit{expression}, \text{`)'}\right]\right),$\\$\qquad\qquad\mathit{id},$\\$\qquad\qquad\mathit{number}]))$	\\
$\mathrm{p}(\text{`'},\mathit{op},\mathrm{choice}([\text{`+'},$\\$\qquad\qquad\text{`-'},$\\$\qquad\qquad\text{`=='}]))$	\\
\hline\end{tabular}\end{center}

\section{Normalizations}
{\footnotesize\begin{itemize}
\item \textbf{abstractize-concretize}\\$\mathrm{p}\left(\text{`'},\mathit{fun},\mathrm{seq}\left(\left[\mathit{id}, \plus \left(\mathit{id}\right), \fbox{$\text{`='}$}, \mathit{expression}, \mathit{newline}\right]\right)\right)$
\item \textbf{abstractize-concretize}\\$\mathrm{p}\left(\text{`'},\mathit{op},\mathrm{choice}\left(\left[\fbox{$\text{`+'}$}, \fbox{$\text{`-'}$}, \fbox{$\text{`=='}$}\right]\right)\right)$
\item \textbf{abstractize-concretize}\\$\mathrm{p}\left(\text{`'},\mathit{expression},\mathrm{choice}\left(\left[\mathrm{seq}\left(\left[\mathit{expression}, \mathit{op}, \mathit{expression}\right]\right), \mathrm{seq}\left(\left[\mathit{id}, \plus \left(\mathit{expression}\right)\right]\right), \mathrm{seq}\left(\left[\fbox{$\text{`if'}$}, \mathit{expression}, \fbox{$\text{`then'}$}, \mathit{expression}, \fbox{$\text{`else'}$}, \mathit{expression}\right]\right), \mathrm{seq}\left(\left[\fbox{$\text{`('}$}, \mathit{expression}, \fbox{$\text{`)'}$}\right]\right), \mathit{id}, \mathit{number}\right]\right)\right)$
\item \textbf{vertical-horizontal}  in $\mathit{expression}$
\item \textbf{undefine-define}\\$\mathrm{p}\left(\text{`'},\mathit{op},\varepsilon\right)$
\item \textbf{abridge-detour}\\$\mathrm{p}\left(\text{`'},\mathit{expression},\mathit{expression}\right)$
\item \textbf{extract-inline}  in $\mathit{expression}$\\$\mathrm{p}\left(\text{`'},\mathit{expression_1},\mathrm{seq}\left(\left[\mathit{expression}, \mathit{op}, \mathit{expression}\right]\right)\right)$
\item \textbf{extract-inline}  in $\mathit{expression}$\\$\mathrm{p}\left(\text{`'},\mathit{expression_2},\mathrm{seq}\left(\left[\mathit{id}, \plus \left(\mathit{expression}\right)\right]\right)\right)$
\item \textbf{extract-inline}  in $\mathit{expression}$\\$\mathrm{p}\left(\text{`'},\mathit{expression_3},\mathrm{seq}\left(\left[\mathit{expression}, \mathit{expression}, \mathit{expression}\right]\right)\right)$
\end{itemize}}

\section{Grammar in ANF}

\footnotesize\begin{center}\begin{tabular}{|l|c|}\hline
\multicolumn{1}{|>{\columncolor[gray]{.9}}c|}{\footnotesize \textbf{Production rule}} &
\multicolumn{1}{>{\columncolor[gray]{.9}}c|}{\footnotesize \textbf{Production signature}}
\\\hline
$\mathrm{p}\left(\text{`'},\mathit{program},\plus \left(\mathit{fun}\right)\right)$	&	$\{ \langle \mathit{fun}, {+}\rangle\}$\\
$\mathrm{p}\left(\text{`'},\mathit{fun},\mathrm{seq}\left(\left[\mathit{id}, \plus \left(\mathit{id}\right), \mathit{expression}, \mathit{newline}\right]\right)\right)$	&	$\{ \langle \mathit{newline}, 1\rangle, \langle \mathit{id}, 1{+}\rangle, \langle \mathit{expression}, 1\rangle\}$\\
$\mathrm{p}\left(\text{`'},\mathit{expression},\mathit{expression_1}\right)$	&	$\{ \langle \mathit{expression_1}, 1\rangle\}$\\
$\mathrm{p}\left(\text{`'},\mathit{expression},\mathit{expression_2}\right)$	&	$\{ \langle \mathit{expression_2}, 1\rangle\}$\\
$\mathrm{p}\left(\text{`'},\mathit{expression},\mathit{expression_3}\right)$	&	$\{ \langle \mathit{expression_3}, 1\rangle\}$\\
$\mathrm{p}\left(\text{`'},\mathit{expression},\mathit{id}\right)$	&	$\{ \langle \mathit{id}, 1\rangle\}$\\
$\mathrm{p}\left(\text{`'},\mathit{expression},\mathit{number}\right)$	&	$\{ \langle \mathit{number}, 1\rangle\}$\\
$\mathrm{p}\left(\text{`'},\mathit{expression_1},\mathrm{seq}\left(\left[\mathit{expression}, \mathit{op}, \mathit{expression}\right]\right)\right)$	&	$\{ \langle \mathit{op}, 1\rangle, \langle \mathit{expression}, 11\rangle\}$\\
$\mathrm{p}\left(\text{`'},\mathit{expression_2},\mathrm{seq}\left(\left[\mathit{id}, \plus \left(\mathit{expression}\right)\right]\right)\right)$	&	$\{ \langle \mathit{expression}, {+}\rangle, \langle \mathit{id}, 1\rangle\}$\\
$\mathrm{p}\left(\text{`'},\mathit{expression_3},\mathrm{seq}\left(\left[\mathit{expression}, \mathit{expression}, \mathit{expression}\right]\right)\right)$	&	$\{ \langle \mathit{expression}, 111\rangle\}$\\
\hline\end{tabular}\end{center}

\section{Nominal resolution}

Production rules are matched as follows (ANF on the left, master grammar on the right):
\begin{eqnarray*}
\mathrm{p}\left(\text{`'},\mathit{program},\plus \left(\mathit{fun}\right)\right) & \bumpeq & \mathrm{p}\left(\text{`'},\mathit{program},\plus \left(\mathit{function}\right)\right) \\
\mathrm{p}\left(\text{`'},\mathit{fun},\mathrm{seq}\left(\left[\mathit{id}, \plus \left(\mathit{id}\right), \mathit{expression}, \mathit{newline}\right]\right)\right) & \Bumpeq & \mathrm{p}\left(\text{`'},\mathit{function},\mathrm{seq}\left(\left[str, \plus \left(str\right), \mathit{expression}\right]\right)\right) \\
\mathrm{p}\left(\text{`'},\mathit{expression},\mathit{expression_1}\right) & \bumpeq & \mathrm{p}\left(\text{`'},\mathit{expression},\mathit{binary}\right) \\
\mathrm{p}\left(\text{`'},\mathit{expression},\mathit{expression_2}\right) & \bumpeq & \mathrm{p}\left(\text{`'},\mathit{expression},\mathit{apply}\right) \\
\mathrm{p}\left(\text{`'},\mathit{expression},\mathit{expression_3}\right) & \bumpeq & \mathrm{p}\left(\text{`'},\mathit{expression},\mathit{conditional}\right) \\
\mathrm{p}\left(\text{`'},\mathit{expression},\mathit{id}\right) & \bumpeq & \mathrm{p}\left(\text{`'},\mathit{expression},str\right) \\
\mathrm{p}\left(\text{`'},\mathit{expression},\mathit{number}\right) & \bumpeq & \mathrm{p}\left(\text{`'},\mathit{expression},int\right) \\
\mathrm{p}\left(\text{`'},\mathit{expression_1},\mathrm{seq}\left(\left[\mathit{expression}, \mathit{op}, \mathit{expression}\right]\right)\right) & \bumpeq & \mathrm{p}\left(\text{`'},\mathit{binary},\mathrm{seq}\left(\left[\mathit{expression}, \mathit{operator}, \mathit{expression}\right]\right)\right) \\
\mathrm{p}\left(\text{`'},\mathit{expression_2},\mathrm{seq}\left(\left[\mathit{id}, \plus \left(\mathit{expression}\right)\right]\right)\right) & \bumpeq & \mathrm{p}\left(\text{`'},\mathit{apply},\mathrm{seq}\left(\left[str, \plus \left(\mathit{expression}\right)\right]\right)\right) \\
\mathrm{p}\left(\text{`'},\mathit{expression_3},\mathrm{seq}\left(\left[\mathit{expression}, \mathit{expression}, \mathit{expression}\right]\right)\right) & \bumpeq & \mathrm{p}\left(\text{`'},\mathit{conditional},\mathrm{seq}\left(\left[\mathit{expression}, \mathit{expression}, \mathit{expression}\right]\right)\right) \\
\end{eqnarray*}
This yields the following nominal mapping:
\begin{align*}\mathit{txl} \:\diamond\: \mathit{master} =\:& \{\langle \mathit{program},\mathit{program}\rangle,\\
 & \langle \mathit{expression_2},\mathit{apply}\rangle,\\
 & \langle \mathit{fun},\mathit{function}\rangle,\\
 & \langle \mathit{expression},\mathit{expression}\rangle,\\
 & \langle \mathit{id},str\rangle,\\
 & \langle \mathit{expression_1},\mathit{binary}\rangle,\\
 & \langle \mathit{op},\mathit{operator}\rangle,\\
 & \langle \mathit{number},int\rangle,\\
 & \langle \mathit{newline},\omega\rangle,\\
 & \langle \mathit{expression_3},\mathit{conditional}\rangle\}\end{align*}
 Which is exercised with these grammar transformation steps:

{\footnotesize\begin{itemize}
\item \textbf{renameN-renameN} $\mathit{expression_2}$ to $\mathit{apply}$
\item \textbf{renameN-renameN} $\mathit{fun}$ to $\mathit{function}$
\item \textbf{renameN-renameN} $\mathit{id}$ to $str$
\item \textbf{renameN-renameN} $\mathit{expression_1}$ to $\mathit{binary}$
\item \textbf{renameN-renameN} $\mathit{op}$ to $\mathit{operator}$
\item \textbf{renameN-renameN} $\mathit{number}$ to $int$
\item \textbf{renameN-renameN} $\mathit{expression_3}$ to $\mathit{conditional}$
\end{itemize}}

\section{Structural resolution}
{\footnotesize\begin{itemize}
\item \textbf{project-inject}\\$\mathrm{p}\left(\text{`'},\mathit{function},\mathrm{seq}\left(\left[str, \plus \left(str\right), \mathit{expression}, \fbox{$\mathit{newline}$}\right]\right)\right)$
\end{itemize}}

\chapter{XML Schema}

 Source name: \textbf{xsd}

\section{Source grammar}

\begin{itemize}\item Source artifact: \href{http://github.com/grammarware/slps/blob/master/topics/fl/xsd/fl.xsd}{topics/fl/xsd/fl.xsd}\item Grammar extractor: \href{http://github.com/grammarware/slps/blob/master/shared/prolog/xsd2bgf.pro}{shared/prolog/xsd2bgf.pro}\end{itemize}

\footnotesize\begin{center}\begin{tabular}{|l|}\hline
\multicolumn{1}{|>{\columncolor[gray]{.9}}c|}{\footnotesize \textbf{Production rules}}
\\\hline
$\mathrm{p}(\text{`'},\mathit{Program},\plus \left(\mathrm{sel}\left(\text{`function'},\mathit{Function}\right)\right))$	\\
$\mathrm{p}(\text{`'},\mathit{Fragment},\mathit{Expr})$	\\
$\mathrm{p}(\text{`'},\mathit{Function},\mathrm{seq}\left(\left[\mathrm{sel}\left(\text{`name'},str\right), \plus \left(\mathrm{sel}\left(\text{`arg'},str\right)\right), \mathrm{sel}\left(\text{`rhs'},\mathit{Expr}\right)\right]\right))$	\\
$\mathrm{p}(\text{`'},\mathit{Expr},\mathrm{choice}([\mathit{Literal},$\\$\qquad\qquad\mathit{Argument},$\\$\qquad\qquad\mathit{Binary},$\\$\qquad\qquad\mathit{IfThenElse},$\\$\qquad\qquad\mathit{Apply}]))$	\\
$\mathrm{p}(\text{`'},\mathit{Literal},\mathrm{sel}\left(\text{`info'},int\right))$	\\
$\mathrm{p}(\text{`'},\mathit{Argument},\mathrm{sel}\left(\text{`name'},str\right))$	\\
$\mathrm{p}(\text{`'},\mathit{Binary},\mathrm{seq}\left(\left[\mathrm{sel}\left(\text{`ops'},\mathit{Ops}\right), \mathrm{sel}\left(\text{`left'},\mathit{Expr}\right), \mathrm{sel}\left(\text{`right'},\mathit{Expr}\right)\right]\right))$	\\
$\mathrm{p}(\text{`'},\mathit{Ops},\mathrm{choice}([\mathrm{sel}\left(\text{`Equal'},\varepsilon\right),$\\$\qquad\qquad\mathrm{sel}\left(\text{`Plus'},\varepsilon\right),$\\$\qquad\qquad\mathrm{sel}\left(\text{`Minus'},\varepsilon\right)]))$	\\
$\mathrm{p}(\text{`'},\mathit{IfThenElse},\mathrm{seq}\left(\left[\mathrm{sel}\left(\text{`ifExpr'},\mathit{Expr}\right), \mathrm{sel}\left(\text{`thenExpr'},\mathit{Expr}\right), \mathrm{sel}\left(\text{`elseExpr'},\mathit{Expr}\right)\right]\right))$	\\
$\mathrm{p}(\text{`'},\mathit{Apply},\mathrm{seq}\left(\left[\mathrm{sel}\left(\text{`name'},str\right), \plus \left(\mathrm{sel}\left(\text{`arg'},\mathit{Expr}\right)\right)\right]\right))$	\\
\hline\end{tabular}\end{center}

\section{Normalizations}
{\footnotesize\begin{itemize}
\item \textbf{unlabel-designate}\\$\mathrm{p}\left(\fbox{\text{`info'}},\mathit{Literal},int\right)$
\item \textbf{unlabel-designate}\\$\mathrm{p}\left(\fbox{\text{`name'}},\mathit{Argument},str\right)$
\item \textbf{anonymize-deanonymize}\\$\mathrm{p}\left(\text{`'},\mathit{Apply},\mathrm{seq}\left(\left[\fbox{$\mathrm{sel}\left(\text{`name'},str\right)$}, \plus \left(\fbox{$\mathrm{sel}\left(\text{`arg'},\mathit{Expr}\right)$}\right)\right]\right)\right)$
\item \textbf{anonymize-deanonymize}\\$\mathrm{p}\left(\text{`'},\mathit{Function},\mathrm{seq}\left(\left[\fbox{$\mathrm{sel}\left(\text{`name'},str\right)$}, \plus \left(\fbox{$\mathrm{sel}\left(\text{`arg'},str\right)$}\right), \fbox{$\mathrm{sel}\left(\text{`rhs'},\mathit{Expr}\right)$}\right]\right)\right)$
\item \textbf{anonymize-deanonymize}\\$\mathrm{p}\left(\text{`'},\mathit{IfThenElse},\mathrm{seq}\left(\left[\fbox{$\mathrm{sel}\left(\text{`ifExpr'},\mathit{Expr}\right)$}, \fbox{$\mathrm{sel}\left(\text{`thenExpr'},\mathit{Expr}\right)$}, \fbox{$\mathrm{sel}\left(\text{`elseExpr'},\mathit{Expr}\right)$}\right]\right)\right)$
\item \textbf{anonymize-deanonymize}\\$\mathrm{p}\left(\text{`'},\mathit{Program},\plus \left(\fbox{$\mathrm{sel}\left(\text{`function'},\mathit{Function}\right)$}\right)\right)$
\item \textbf{anonymize-deanonymize}\\$\mathrm{p}\left(\text{`'},\mathit{Ops},\mathrm{choice}\left(\left[\fbox{$\mathrm{sel}\left(\text{`Equal'},\varepsilon\right)$}, \fbox{$\mathrm{sel}\left(\text{`Plus'},\varepsilon\right)$}, \fbox{$\mathrm{sel}\left(\text{`Minus'},\varepsilon\right)$}\right]\right)\right)$
\item \textbf{anonymize-deanonymize}\\$\mathrm{p}\left(\text{`'},\mathit{Binary},\mathrm{seq}\left(\left[\fbox{$\mathrm{sel}\left(\text{`ops'},\mathit{Ops}\right)$}, \fbox{$\mathrm{sel}\left(\text{`left'},\mathit{Expr}\right)$}, \fbox{$\mathrm{sel}\left(\text{`right'},\mathit{Expr}\right)$}\right]\right)\right)$
\item \textbf{vertical-horizontal}  in $\mathit{Expr}$
\item \textbf{undefine-define}\\$\mathrm{p}\left(\text{`'},\mathit{Ops},\varepsilon\right)$
\item \textbf{unchain-chain}\\$\mathrm{p}\left(\text{`'},\mathit{Expr},\mathit{Literal}\right)$
\item \textbf{unchain-chain}\\$\mathrm{p}\left(\text{`'},\mathit{Expr},\mathit{Argument}\right)$
\item \textbf{unchain-chain}\\$\mathrm{p}\left(\text{`'},\mathit{Expr},\mathit{Binary}\right)$
\item \textbf{unchain-chain}\\$\mathrm{p}\left(\text{`'},\mathit{Expr},\mathit{IfThenElse}\right)$
\item \textbf{unchain-chain}\\$\mathrm{p}\left(\text{`'},\mathit{Expr},\mathit{Apply}\right)$
\item \textbf{unlabel-designate}\\$\mathrm{p}\left(\fbox{\text{`Literal'}},\mathit{Expr},int\right)$
\item \textbf{unlabel-designate}\\$\mathrm{p}\left(\fbox{\text{`Argument'}},\mathit{Expr},str\right)$
\item \textbf{unlabel-designate}\\$\mathrm{p}\left(\fbox{\text{`Binary'}},\mathit{Expr},\mathrm{seq}\left(\left[\mathit{Ops}, \mathit{Expr}, \mathit{Expr}\right]\right)\right)$
\item \textbf{unlabel-designate}\\$\mathrm{p}\left(\fbox{\text{`IfThenElse'}},\mathit{Expr},\mathrm{seq}\left(\left[\mathit{Expr}, \mathit{Expr}, \mathit{Expr}\right]\right)\right)$
\item \textbf{unlabel-designate}\\$\mathrm{p}\left(\fbox{\text{`Apply'}},\mathit{Expr},\mathrm{seq}\left(\left[str, \plus \left(\mathit{Expr}\right)\right]\right)\right)$
\item \textbf{extract-inline}  in $\mathit{Expr}$\\$\mathrm{p}\left(\text{`'},\mathit{Expr_1},\mathrm{seq}\left(\left[\mathit{Ops}, \mathit{Expr}, \mathit{Expr}\right]\right)\right)$
\item \textbf{extract-inline}  in $\mathit{Expr}$\\$\mathrm{p}\left(\text{`'},\mathit{Expr_2},\mathrm{seq}\left(\left[\mathit{Expr}, \mathit{Expr}, \mathit{Expr}\right]\right)\right)$
\item \textbf{extract-inline}  in $\mathit{Expr}$\\$\mathrm{p}\left(\text{`'},\mathit{Expr_3},\mathrm{seq}\left(\left[str, \plus \left(\mathit{Expr}\right)\right]\right)\right)$
\end{itemize}}

\section{Grammar in ANF}

\footnotesize\begin{center}\begin{tabular}{|l|c|}\hline
\multicolumn{1}{|>{\columncolor[gray]{.9}}c|}{\footnotesize \textbf{Production rule}} &
\multicolumn{1}{>{\columncolor[gray]{.9}}c|}{\footnotesize \textbf{Production signature}}
\\\hline
$\mathrm{p}\left(\text{`'},\mathit{Program},\plus \left(\mathit{Function}\right)\right)$	&	$\{ \langle \mathit{Function}, {+}\rangle\}$\\
$\mathrm{p}\left(\text{`'},\mathit{Fragment},\mathit{Expr}\right)$	&	$\{ \langle \mathit{Expr}, 1\rangle\}$\\
$\mathrm{p}\left(\text{`'},\mathit{Function},\mathrm{seq}\left(\left[str, \plus \left(str\right), \mathit{Expr}\right]\right)\right)$	&	$\{ \langle str, 1{+}\rangle, \langle \mathit{Expr}, 1\rangle\}$\\
$\mathrm{p}\left(\text{`'},\mathit{Expr},int\right)$	&	$\{ \langle int, 1\rangle\}$\\
$\mathrm{p}\left(\text{`'},\mathit{Expr},str\right)$	&	$\{ \langle str, 1\rangle\}$\\
$\mathrm{p}\left(\text{`'},\mathit{Expr},\mathit{Expr_1}\right)$	&	$\{ \langle \mathit{Expr_1}, 1\rangle\}$\\
$\mathrm{p}\left(\text{`'},\mathit{Expr},\mathit{Expr_2}\right)$	&	$\{ \langle \mathit{Expr_2}, 1\rangle\}$\\
$\mathrm{p}\left(\text{`'},\mathit{Expr},\mathit{Expr_3}\right)$	&	$\{ \langle \mathit{Expr_3}, 1\rangle\}$\\
$\mathrm{p}\left(\text{`'},\mathit{Expr_1},\mathrm{seq}\left(\left[\mathit{Ops}, \mathit{Expr}, \mathit{Expr}\right]\right)\right)$	&	$\{ \langle \mathit{Ops}, 1\rangle, \langle \mathit{Expr}, 11\rangle\}$\\
$\mathrm{p}\left(\text{`'},\mathit{Expr_2},\mathrm{seq}\left(\left[\mathit{Expr}, \mathit{Expr}, \mathit{Expr}\right]\right)\right)$	&	$\{ \langle \mathit{Expr}, 111\rangle\}$\\
$\mathrm{p}\left(\text{`'},\mathit{Expr_3},\mathrm{seq}\left(\left[str, \plus \left(\mathit{Expr}\right)\right]\right)\right)$	&	$\{ \langle str, 1\rangle, \langle \mathit{Expr}, {+}\rangle\}$\\
\hline\end{tabular}\end{center}

\section{Nominal resolution}

Production rules are matched as follows (ANF on the left, master grammar on the right):
\begin{eqnarray*}
\mathrm{p}\left(\text{`'},\mathit{Program},\plus \left(\mathit{Function}\right)\right) & \bumpeq & \mathrm{p}\left(\text{`'},\mathit{program},\plus \left(\mathit{function}\right)\right) \\
\mathrm{p}\left(\text{`'},\mathit{Fragment},\mathit{Expr}\right) &  & \varnothing \\
\mathrm{p}\left(\text{`'},\mathit{Function},\mathrm{seq}\left(\left[str, \plus \left(str\right), \mathit{Expr}\right]\right)\right) & \bumpeq & \mathrm{p}\left(\text{`'},\mathit{function},\mathrm{seq}\left(\left[str, \plus \left(str\right), \mathit{expression}\right]\right)\right) \\
\mathrm{p}\left(\text{`'},\mathit{Expr},int\right) & \bumpeq & \mathrm{p}\left(\text{`'},\mathit{expression},int\right) \\
\mathrm{p}\left(\text{`'},\mathit{Expr},str\right) & \bumpeq & \mathrm{p}\left(\text{`'},\mathit{expression},str\right) \\
\mathrm{p}\left(\text{`'},\mathit{Expr},\mathit{Expr_1}\right) & \bumpeq & \mathrm{p}\left(\text{`'},\mathit{expression},\mathit{binary}\right) \\
\mathrm{p}\left(\text{`'},\mathit{Expr},\mathit{Expr_2}\right) & \bumpeq & \mathrm{p}\left(\text{`'},\mathit{expression},\mathit{conditional}\right) \\
\mathrm{p}\left(\text{`'},\mathit{Expr},\mathit{Expr_3}\right) & \bumpeq & \mathrm{p}\left(\text{`'},\mathit{expression},\mathit{apply}\right) \\
\mathrm{p}\left(\text{`'},\mathit{Expr_1},\mathrm{seq}\left(\left[\mathit{Ops}, \mathit{Expr}, \mathit{Expr}\right]\right)\right) & \Bumpeq & \mathrm{p}\left(\text{`'},\mathit{binary},\mathrm{seq}\left(\left[\mathit{expression}, \mathit{operator}, \mathit{expression}\right]\right)\right) \\
\mathrm{p}\left(\text{`'},\mathit{Expr_2},\mathrm{seq}\left(\left[\mathit{Expr}, \mathit{Expr}, \mathit{Expr}\right]\right)\right) & \bumpeq & \mathrm{p}\left(\text{`'},\mathit{conditional},\mathrm{seq}\left(\left[\mathit{expression}, \mathit{expression}, \mathit{expression}\right]\right)\right) \\
\mathrm{p}\left(\text{`'},\mathit{Expr_3},\mathrm{seq}\left(\left[str, \plus \left(\mathit{Expr}\right)\right]\right)\right) & \bumpeq & \mathrm{p}\left(\text{`'},\mathit{apply},\mathrm{seq}\left(\left[str, \plus \left(\mathit{expression}\right)\right]\right)\right) \\
\end{eqnarray*}
This yields the following nominal mapping:
\begin{align*}\mathit{xsd} \:\diamond\: \mathit{master} =\:& \{\langle \mathit{Expr_1},\mathit{binary}\rangle,\\
 & \langle str,str\rangle,\\
 & \langle int,int\rangle,\\
 & \langle \mathit{Expr_2},\mathit{conditional}\rangle,\\
 & \langle \mathit{Function},\mathit{function}\rangle,\\
 & \langle \mathit{Program},\mathit{program}\rangle,\\
 & \langle \mathit{Expr_3},\mathit{apply}\rangle,\\
 & \langle \mathit{Expr},\mathit{expression}\rangle,\\
 & \langle \mathit{Ops},\mathit{operator}\rangle\}\end{align*}
 Which is exercised with these grammar transformation steps:

{\footnotesize\begin{itemize}
\item \textbf{renameN-renameN} $\mathit{Expr_1}$ to $\mathit{binary}$
\item \textbf{renameN-renameN} $\mathit{Expr_2}$ to $\mathit{conditional}$
\item \textbf{renameN-renameN} $\mathit{Function}$ to $\mathit{function}$
\item \textbf{renameN-renameN} $\mathit{Program}$ to $\mathit{program}$
\item \textbf{renameN-renameN} $\mathit{Expr_3}$ to $\mathit{apply}$
\item \textbf{renameN-renameN} $\mathit{Expr}$ to $\mathit{expression}$
\item \textbf{renameN-renameN} $\mathit{Ops}$ to $\mathit{operator}$
\end{itemize}}

\section{Structural resolution}
{\footnotesize\begin{itemize}
\item \textbf{reroot-reroot} $\left[\mathit{program}, \mathit{Fragment}\right]$ to $\left[\mathit{program}\right]$
\item \textbf{eliminate-introduce}\\$\mathrm{p}\left(\text{`'},\mathit{Fragment},\mathit{expression}\right)$
\item \textbf{permute-permute}\\$\mathrm{p}\left(\text{`'},\mathit{binary},\mathrm{seq}\left(\left[\mathit{operator}, \mathit{expression}, \mathit{expression}\right]\right)\right)$\\$\mathrm{p}\left(\text{`'},\mathit{binary},\mathrm{seq}\left(\left[\mathit{expression}, \mathit{operator}, \mathit{expression}\right]\right)\right)$
\end{itemize}}

\newpage\bibliographystyle{abbrv}
\bibliography{paper}

\begin{thebibliography}{10}

\bibitem{BJV04}
J.~B{\' e}zivin, F.~Jouault, and P.~Valduriez.
\newblock {On the Need for Megamodels}.
\newblock {\em OOPSLA \& GPCE, Workshop on best MDSD practices}, 2004.
\newblock \\Publicly available via
  \url{http://www.softmetaware.com/oopsla2004/bezivin-megamodel.pdf}.

\bibitem{DeanCMS02}
T.~R. Dean, J.~R. Cordy, A.~J. Malton, and K.~A. Schneider.
\newblock {Grammar Programming in TXL}.
\newblock In {\em {Proceedings of SCAM 2002}}. IEEE, 2002.
\newblock \\Publicly available via
  \url{http://plg1.uwaterloo.ca/~ajmalton/ajmalton/Papers/SCAM02_GP.pdf}.

\bibitem{EMF}
{Eclipse}.
\newblock {Eclipse Modeling Framework Project (EMF 2.4)}, 2008.
\newblock \url{http://www.eclipse.org/modeling/emf/}.

\bibitem{EKV09}
G.~Economopoulos, P.~Klint, and J.~J. Vinju.
\newblock Faster scannerless {GLR} parsing.
\newblock In O.~de~Moor and M.~I. Schwartzbach, editors, {\em Proceedings of CC
  2009}, volume 5501 of {\em LNCS}, pages 126--141. Springer, 2009.
\newblock \\Publicly available via
  \url{http://oai.cwi.nl/oai/asset/15095/15095B.pdf}.

\bibitem{MegaL}
J.-M. Favre, R.~L{\"a}mmel, and A.~Varanovich.
\newblock {Modeling the Linguistic Architecture of Software Products}.
\newblock In {\em {Proceedings of MODELS 2012}}, LNCS. Springer, 2012.
\newblock \\Publicly available via \url{http://softlang.uni-koblenz.de/mega}.

\bibitem{FNG04}
J.-M. Favre and T.~NGuyen.
\newblock {Towards a Megamodel to Model Software Evolution through
  Transformations}.
\newblock In {\em Proceedings of SETra}, volume 127 of {\em ENTCS}, 2004.
\newblock \\Publicly available via
  \url{http://adele.imag.fr/Les.Publications/intConferences/SETRAa2004Fav.pdf}.

\bibitem{JSR31}
J.~Fialli and S.~Vajjhala.
\newblock {\em {Java Specification Request 31: XML Data Binding
  Specification}}, 1999.
\newblock \\Publicly available via \url{http://jcp.org/en/jsr/detail?id=031}.

\bibitem{W3C-XSD}
S.~Gao, C.~M. Sperberg-McQueen, and H.~S. Thompson.
\newblock {W3C XML Schema Definition Language (XSD) 1.1 Part 1: Structures}.
\newblock {\em {W3C Recommendation}}, Apr. 2012.
\newblock \\Publicly available via
  \url{http://www.w3.org/TR/2012/REC-xmlschema11-1-20120405}.

\bibitem{Klint93}
P.~Klint.
\newblock {A Meta-Environment for Generating Programming Environments}.
\newblock {\em ACM TOSEM}, 2(2):176--201, 1993.
\newblock \\Publicly available via \url{http://dare.uva.nl/document/28101}.

\bibitem{RascalTutor}
P.~Klint et~al.
\newblock {\em Rascal Tutor}.
\newblock SWAT, CWI, 2012.
\newblock \url{http://tutor.rascal-mpl.org}.

\bibitem{Rascal}
P.~Klint, T.~van~der Storm, and J.~Vinju.
\newblock {EASY Meta-programming with Rascal}.
\newblock In J.~M. Fernandes, R.~L{\"a}mmel, J.~Visser, and J.~Saraiva,
  editors, {\em {Post-proceedings of GTTSE 2009}}, volume 6491 of {\em LNCS},
  pages 222--289. Springer-Verlag, January 2011.
\newblock \\Publicly available via
  \url{http://homepages.cwi.nl/~paulk/publications/rascal-gttse-final.pdf}.

\bibitem{Convergence2009}
R.~L{\"a}mmel and V.~Zaytsev.
\newblock {An Introduction to Grammar Convergence}.
\newblock In M.~Leuschel and H.~Wehrheim, editors, {\em {Proceedings of iFM
  2009}}, volume 5423 of {\em LNCS}, pages 246--260. Springer-Verlag, February
  2009.
\newblock \\Publicly available via
  \url{http://grammarware.net/writes#Convergence2009}.

\bibitem{McGuire2007}
P.~McGuire.
\newblock {\em {Getting Started with Pyparsing}}.
\newblock O'Reilly, first edition, 2007.

\bibitem{MOF}
Object Management Group.
\newblock {\em {Meta-Object Facility (MOF$^\textrm{\tiny TM}$) Core
  Specification}}, 2.0 edition, Jan. 2006.
\newblock \\Publicly available via \url{http://www.omg.org/spec/MOF/2.0}.

\bibitem{ANTLR}
T.~Parr.
\newblock {ANTLR---ANother Tool for Language Recognition}, 2008.
\newblock \url{http://antlr.org}.

\bibitem{DCG}
F.~C.~N. Pereira and D.~H.~D. Warren.
\newblock {Definite Clause Grammars for Language Analysis --- A Survey of the
  Formalism and a Comparison with Augmented Transition Networks}.
\newblock {\em Artificial Intelligence}, 13:231--278, 1980.
\newblock \\Publicly available via
  \url{http://cgi.di.uoa.gr/~takis/pereira-warren.pdf}.

\bibitem{Visser97}
E.~Visser.
\newblock {Scannerless Generalized-{LR} Parsing}.
\newblock Technical Report P9707, Programming Research Group, Universiteit van
  Amsterdam, July 1997.
\newblock \\Publicly available via
  \url{http://www.science.uva.nl/pub/programming-research/reports/1997/P9707.ps.Z}.

\bibitem{Guided2013}
V.~Zaytsev.
\newblock {Guided Grammar Convergence}.
\newblock Submitted to POPL 2013. Pending notification. June 2012. \\Publicly
  available via \url{http://grammarware.net/writes#Guided2013}.

\bibitem{Renarration2012}
V.~Zaytsev.
\newblock {Renarrating Linguistic Architecture: A Case Study}.
\newblock Submitted to MPM 2012. Pending notification. July 2012. \\Publicly
  available via \url{http://grammarware.net/writes#Renarration2012}.

\bibitem{Metasyntactically2012}
V.~Zaytsev.
\newblock {Language Evolution, Metasyntactically}.
\newblock In {\em Proceedings of BX 2012}, volume~49 of {\em EC-EASST}. EASST,
  2012.
\newblock \\Publicly available via
  \url{http://grammarware.net/writes#Metasyntactically2012}.

\bibitem{SLPS}
V.~Zaytsev, R.~L{\"a}mmel, T.~van~der Storm, L.~Renggli, and G.~Wachsmuth.
\newblock {Software Language Processing Suite\footnote{The authors are given
  according to the statistics at
  \url{http://github.com/grammarware/slps/graphs/contributors}.}}, 2008--2012.
\newblock \url{http://grammarware.github.com}.

\end{thebibliography}

\end{document}